\documentclass[11pt,aps,floats,showpacs,amssymb,tightenlines,nofootinbib]{revtex4}
\usepackage{color, colortbl}
\usepackage{graphicx}
\usepackage{amsfonts}
\usepackage{amscd}
\usepackage{epsfig}
\usepackage{amssymb}
\usepackage{tabularx}
\usepackage{fancyhdr}
\usepackage{amsmath,amssymb}
\usepackage[T1]{fontenc}
\usepackage{calligra}
\usepackage[all]{xy}

\definecolor{Gray}{gray}{0.9}
\definecolor{LightCyan}{rgb}{0.88,1,1}
\begin{document}

\title[]{Evolution of thin shells in D-dimensional General Relativity}
\author{Marcos A. Ramirez$^{1,2}$} \email{mramirez@famaf.unc.edu.ar}%
\author{Daniel Aparicio$^2$}
\affiliation{$^1$Instituto de F\'{i}sica Enrique Gaviola, FaMAF, Universidad Nacional de C\'ordoba, (5000) C\'ordoba, Argentina }
\affiliation{$^2$Instituto de Investigaciones en Energ\'{i}a no Convencional, Facultad de Ciencias Exactas, Universidad Nacional de Salta, (4400) Salta, Argentina}
%\thanks{}%
%\date{}%
% ----------------------------------------------------------------
\begin{abstract}

In this paper we consider singular timelike spherical hypersurfaces embedded in a $D$-dimensional spherically symmetric bulk spacetime which is an electrovacuum solution of Einstein equations with cosmological constant. We analyse the different possibilities regarding the orientation of the gradient of the standard $r$ coordinate in relation to the shell. Then we study the dynamics according to Einstein equations for arbitrary matter satisfying the dominant energy condition. In particular, we thoroughly analyse the asymptotic dynamics for both the small and large-shell-radius limits. We also study the main qualitative aspects of the dynamics of shells made of linear barotropic fluids that satisfy the dominant energy condition. Finally, we prove weak cosmic censorship for this class of solutions.

\end{abstract}

\maketitle
% ----------------------------------------------------------------

\section{Introduction}

%LOOK FOR GRAVITY WAVES THIN SHELL!

Thin shell models are interesting perhaps for two main reasons. The first one is the possibility of a mathematical simplification of the field equations that may allow us to characterize non-vacuum solutions. In this case the thin shell models are not necessarily realistic, they are useful because of their mathematical properties, and they may shed light on general theoretical problems such as the cosmic censorship conjecture \cite{censor} %emission of gravitational waves and the associated back-reaction \cite{emission}, 
or the possibility of transversable wormholes \cite{TSE,Ref9ES}. 
It is then not surprising that they have been used to address different aspects of higher-dimensional gravity theories, not only general relativity in higher-dimensions but also Lovelock gravity and Chern-Simons gravity, just to name a few.  
%In this sense, thin shell solutions of higher-dimensional gravity theories are always interesting because they might help us to understand some fundamental properties of the purportedly actual 4-dimensional scenarios....    

The second reason is the description of physical systems that are the source of a field theory and that can be modelled by neglecting one of their dimensions. In the context of gravity, there exist a number of astrophysical and cosmological systems (\cite{astrophysics}, \cite{phasetransition}) where some part of them can be thought of as a thin shell to a good approximation. A currently relevant scenario where thin shells play a role in this sense are brane-world cosmologies (for a recent review see \cite{Review}). In this scenario the observable universe is a part of a 4-dimensional manifold embedded in a higher dimensional one, usually 5-dimensional. 
%It has been considered in the literature even higher-dimensional models %(look for 6,7,8, jeje ) as well, the majority them stemming from string-theoretical or M-theoretical considerations. 
In contrast to the more usual compactification scenario, in the context of brane-world gravity it makes perfect sense to 
analyse higher-dimensional solutions of Einstein equations because they might represent classical limits of M-theory.     
%Partly as a consequence of the interest that these models have attracted, in the last two decades there have been a great number of works analyzing higher-dimensional solutions of Einstein equations. 

%Taking into account the diversity of scenarios where thin shells in higher dimensions are studied we can safely state that the analysis of general properties of thin shells solutions in arbitrary dimensions are relevant.

In particular, in recent years several different classical solutions involving self-gravitating thin shells in arbitrary dimensions have been studied \cite{CrisostomoOlea,EiroaSimeone,GaoLemos,BJB}. The generalization of different physical systems into arbitrary dimensions, specially if there are for the system at least two different relevant dimensionalities, is interesting not only because of its synthesis capability, but also because of the possibility of achieving a deeper understanding of the interactions that govern the system and the meaning of dimensionality itself. In \cite{GaoLemos}, Gao and Lemos considered a charged dust shell for a spherically symmetric electro-vacuum bulk of arbitrary dimensions. They analysed the dynamics for different parameters of the bulk regions and considered the possibility of a violation of the cosmic censorship conjecture, as for certain bulk parameters the spacetime would display a naked singularity. They thoroughly analysed the parameter space of this scenario and concluded that cosmic censorship holds in all dimensionalities. On the other hand, in \cite{EiroaSimeone}, Eiroa and Simeone analysed a more general situation with a cosmological constant and matter satisfying the weak energy condition. %They put emphasis in the static solutions and their stability for arbitrary matter-energy content and analysed this situation thoroughly. 
% shells made of matter-energy that satisfy the weak energy condition in a spherically symmetric electro-vacuum bulk with cosmological constant. 
Their analysis laid emphasis in the stability of static solutions for arbitrary matter-energy models, but also considered some general aspects of the dynamics for certain particular cases.  
%cite 8,9,10 of Eiroa - Simeone (8,9 higher dimensional whormholes, 10 D-dimensional dust collapse)

Matter models are somewhat arbitrary, and the knowledge of the possible dynamics of the shell regardless of their matter content allow us to constraint our expectations regarding the evolution of these systems. In this spirit we study certain key aspects of the dynamics of thin shells for any matter content satisfying the dominant energy condition embedded in an electro-vacuum bulk spacetime with cosmological constant and arbitrary dimensionality. 
%The spacetime is in all cases spherically symmetric with cosmological constant and it is a solution of the Einstein-Maxwell system for the corresponding dimensionality.
This work generalizes in a sense the results of both \cite{GaoLemos} and \cite{EiroaSimeone}. We analyse the dynamics of these shells in a more systematic way regarding the different matter-energy models, and recover the main results of these two previous works but in a more general setting. In particular, we prove that weak cosmic censorship holds for this more encompassing class of solutions.

\subsubsection*{Outline}

We begin with a description of a general shell in a spherically symmetric bulk spacetime of an arbitrary number of dimensions in Section \ref{sphericalshells}. In Section \ref{sphericalmotion} we analyse the asymptotic behaviour of the dynamics of these shells, provided their matter-energy content satisfy the dominant energy condition. In Section \ref{barotropic} we describe key properties of the dynamics of shells made of linear barotropic fluids for the different scenarios and equations of state. Finally, in Section \ref{final} we summarize the analysis made in the previous Sections and prove weak cosmic censorship for these solutions.

\section{Spherically symmetric thin shells in $D=n+2$ dimensions}
\label{sphericalshells}
We begin with a description of a singular shell embedded in a spherically symmetric electrovacuum spacetime of an arbitrary number of dimensions. 
%The idea of spherical symmetry is originally defined for $n+1$-dimensional riemannian manifolds: there is a group of isometries isomorphic to the $SO(n)$ group.
 %and it is homeomorphic to the exterior of a ball in $\mathds{R}^3$ \cite{Choquet-Bruhat}
%For lorentzian manifolds, we say that there exists a spacelike foliation such that every riemannian slice is spherically symmetric and the lapse and shift are invariant against the action of the group. Charting the space of group orbits with orthogonal coordinates $(x_0,x_1)$, which is always possible, we may write
If this symmetry holds then there is always, at least locally, an orthogonal coordinate chart $(x_0,x_1)$ for the quotient manifold, so the metric can be written as follows

\begin{equation}
\label{spherical}
ds^2=-f(x_0,x_1)dx_0^2+h(x_0,x_1)dx_1^2+r(x_0,x_1)^2d\Omega_n^2 .
\end{equation}

We now define a singular timelike orientable hypersurface $\Sigma$ embedded in the bulk spacetime. Because of the symmetry, the surface can be described by an equation $\Sigma(x^0,x^1)=0$. In a neighbourhood of $\Sigma$, in gaussian coordinates, the metric reads 
\begin{equation}
ds^2=-f(\tau,\eta)d\tau^2+d\eta^2+r(\tau,\eta)^2d\Omega_n^2
\end{equation}
where $\eta=0$ characterises the surface, and $\tau$ is the shell proper time, so $f(\tau,0)=1$.

In the context of thin shells Einstein equations are equivalent to junction conditions on the surface \cite{Israel} that relate the jump of its extrinsic curvature with the effective stress-energy tensor on the shell, the so-called Darmois-Israel junction conditions. The extrinsic curvature in gaussian coordinates can be written,
\begin{equation}
\label{extrinsic}
K^i_j = \mbox{diag} \left[\left.\left.\left.\frac{1}{2}\frac{\partial f}{\partial\eta}\right|_{\eta=0},\frac{1}{r}\frac{\partial r}{\partial\eta}\right|_{\eta=0},..,\frac{1}{r}\frac{\partial r}{\partial\eta}\right|_{\eta=0}\right]
\end{equation}
where latin indexes represent coordinates $(\tau,\theta_1,..,\theta_n)$ on the shell. In these coordinates the intrinsic metric reads,
\begin{equation}
\label{intrinsic}
ds^2_{\Sigma} = -d\tau^2 + R(\tau)^2 d\Omega_n^2
\end{equation}
where $R(\tau)\equiv r(\tau,0)$.

%\begin{eqnarray}
%\label{Isr1}
%-(n^{\alpha}n_{\alpha})\left(\left[K_{ij}-\frac{1}{2}h_{ij} K\right]\right)&=&\kappa S_{ij}
%\label{Isr2}
%S^{ij}[K_{ij}]_{+} &= &-\kappa[T_{\mu\nu}n^{\mu}n^{\nu}] \\
%\label{Isr3}
%S^{j}_{i;j} & = &(n^{\alpha}n_{\alpha})[T_{\mu\nu}n^{\mu}e_i^{\nu}]
%\end{eqnarray}
%where $[a]$ represent the jump, $[a]_+$ represent the sum of the values of $a$ at both sides, latin indexes represent shell coordinates, greek indexes bulk coordinates.

By virtue of a generalized Birkhoff theorem, electrovacuum solutions of Einstein equations with a cosmological constant can always be written in the form \cite{tangherlini} \cite{myersperry},
%CHECK THE BIRKHOFF THEOREM ASSERTION
\begin{equation}
\label{schwarzschild}
ds^2=-F(r)dt^2+F(r)^{-1}dr^2+r^2d\Omega_n^2 ,   \;\;\; where \;\;\; F(r)=1-\frac{2M}{r^{n-1}}+\frac{Q^2}{r^{2(n-1)}}-\frac{2\Lambda}{n(n+1)}r^2 ,
%ds^2=-\left(1-\frac{2M}{r^{n-1}}\right)dt^2+\left(1-\frac{2M}{r^{n-1}}\right)^{-1}dr^2+r^2d\Omega_n^2
\end{equation}
where $M=\kappa_D m / (n\Omega_n)$ ($m$ is the gravitational contribution to Misner-Sharp energy as defined in any group orbit, $\Omega_n$ is the area of a $n$-sphere of unit radius), and $Q^2=2q^2/(n(n-1))$ ($q$ is analogously the electric charge defined in any group orbit).

%From now on, we assume $F(r)>0$, unless otherwise stated, since we are not interested in collapsing situations. 
With these expressions, we can write the extrinsic curvature on a given {\it side} of the shell in terms of $R(\tau)$ and the parameters $(M,Q)$ that characterise the bulk spacetime there    
%For that we only need to express the first derivatives at the shell of the transformation $(t,r)\\leftrightarrow (\tau,\eta)$
%qu� pasa si r es temporal?  metemos el signo en la expresi�n general?
\begin{equation}
\label{extrinsic2}
K^i_j = \mbox{sign}\left(\left.\frac{\partial r}{\partial \eta}\right|_{\eta=0}\right)\mbox{diag}
\left[\frac{F'(R)+2\ddot{R}}{2\sqrt{\dot{R}^2+F(R)}},\frac{\sqrt{\dot{R}^2+F(R)}}{R},..,\frac{\sqrt{\dot{R}^2+F(R)}}{R}\right].
\end{equation}
We assume that the cosmological constant $\Lambda$ is the same at both sides of the shell, because it is a part of the field equations we are solving.
In this way, giving $(M,Q)$ and specifying whether $r$ increases or decreases with $\eta$, we get an expression for the extrinsic curvature in terms of the function $R(\tau)$.

On the other hand, the matter content of the shell is described by a tensor $S^i_j$ defined on $\Sigma$ such that we can formally write the $D$-dimensional stress-energy tensor as,
\begin{equation}
T^a_b = \delta(\Sigma) S^a_b
\end{equation}
where the tensor $S$ in shell coordinates can be written as follows
\begin{equation}
\label{S}
S^i_j = \mbox{diag} [-\rho(\tau),p(\tau),...,p(\tau)].
\end{equation}
So, as a result of the symmetry imposed, the matter content of the shell can be described as if it were an $n$-dimensional perfect fluid, whose flow lines follow the trajectories of the comoving observers. We can write
\begin{equation}
S^{ij}=p h^{ij} +(\rho + p)u^iu^j,
\end{equation}
where $h_{ij}$ is the intrinsic metric defined in (\ref{intrinsic}) and $u^i=(\partial/\partial\tau)^i$ is the $4$-velocity of the aforementioned comoving observers. If there is not hysteresis, we would be able to write $\rho$ and $p$ as functions of $R$. In that case, conservation of the source would imply
\begin{equation}
\label{conservation}
\frac{d\rho}{dR}+\frac{n(\rho+p)}{R}=0.
\end{equation}
This equation together with an equation of state $f(\rho,p)=0$, provided there is one, settle $\rho(R)$ and $p(R)$. Alternatively, if one gives $\rho(R)$, then $p(R)$ can be derived from (\ref{conservation}). The converse is also true, up to an integration constant. Throughout this paper we will impose the dominant energy condition (DEC) for the matter-energy content of the shell, which can be written as $|p(R)|\leq\rho(R)$.

In this way, the Darmois-Israel junction conditions relate $R(\tau)$, it first two derivatives, and the parameters $(M,Q)$ with the matter functions $\rho(R)$ and $p(R)$. 
Looking at (\ref{extrinsic2}), one can notice that the discontinuity of the extrinsic curvature should be ascribed to a difference between the mass parameters or the charge parameters for the empty regions at both sides of the shell, which we call $(M_I,Q_I)$ and $(M_{II},Q_{II})$, and, eventually, to different signs for $\partial r/\partial \eta$ at both sides.
The signs of these derivatives define the character of the bulk regions that are being glued. Without any loss of generality we choose the $\eta$ coordinate to decrease when going into region $I$ and to increase into region $II$. We define $\xi_{I}=\mbox{sign}\left.\left(\frac{\partial r}{\partial \eta}\right)\right|_{\eta=0^-}$ and $\xi_{II}=\mbox{sign}\left.\left(\frac{\partial r}{\partial \eta}\right)\right|_{\eta=0^+}$. If $\xi_I=+1$ ($-1$), then region $I$ must be interior (exterior), which means that it can be characterised by means of an inequality $r<R(t)$ ($r>R(t)$). Analogously, if $\xi_{II}=+1$ ($-1$), then region $II$ must be exterior (interior). With these definitions, the junction conditions can be written as follows

\begin{eqnarray}
\label{Israelndima}
\frac{n}{R}\left(\xi_{II}\sqrt{\dot{R}^2+F_{II}}-\xi_I\sqrt{\dot{R}^2+F_{I}}\right) &=&- \kappa_D \rho \nonumber \\
& & \\
\label{Israelndimb}
\xi_{II}\frac{F'_{II}+2\ddot{R}}{2\sqrt{\dot{R}^2+F_{II}}} - \xi_I\frac{F'_{I}+2\ddot{R}}{2\sqrt{\dot{R}^2+F_{I}}}+\frac{n-1}{R}\left(\xi_{II}\sqrt{\dot{R}^2+F_{II}}-\xi_I\sqrt{\dot{R}^2+F_{I}}\right)&=& \kappa_D p \nonumber , \\
& &
\end{eqnarray}
where $F_i=1-2M_i/R^{n-1}+Q_i^2/R^{2(n-1)}-2\Lambda R^2/(n(n+1))$. From (\ref{Israelndima}) we can see that the dominant energy condition already imposes some constraints on the possible values of the pair $(\xi_I,\xi_{II})$. If $\rho>0$, as required by DEC, then the combination $(\xi_I,\xi_{II})=(-1,+1)$, which would imply that both regions are exterior (the so-called {\it wormhole} orientation), is not possible. As a consequence of this, one of the regions must be interior, and there are essentially two possible scenarios regarding the orientation of the bulk: both regions are interior or one of them is exterior (this last possibility is the so-called {\it standard orientation}).    
In particular, in the case of an exterior region $II$, a positive effective energy density $\rho$ would imply $M_{II}>M_{I}$ or $|Q_I|>|Q_{II}|$ or both.

Also from (\ref{Israelndima}) we can obtain an equation of motion that results independent of the $\xi_i$ and reads
\begin{equation}
\label{eqnmotion1}
\dot{R}^2+ V(R) = 0 \; \; \;\; , \;\; \; \; V(R) = \frac{F_I+F_{II}}{2} - \left(\frac{n(F_I-F_{II})}{2\kappa_D \rho R}\right)^2 -\left(\frac{\kappa_D \rho R}{2n}\right)^2.
\end{equation}
Replacing the functions $F_i$ we obtain
\begin{equation}
\label{potential}
V(R)= 1 -\frac{2\Lambda R^2}{n(n+1)}- \frac{M_I + M_{II}}{R^{n-1}} + \frac{Q_I^2+Q_{II}^2}{2R^{2(n-1)}} - \left(\frac{n(M_{II}-M_{I})}{\kappa_D \rho R^{n}}+\frac{n(Q_I^2-Q_{II}^2)}{2\kappa_D\rho R^{2n-1}}\right)^2-\frac{\kappa_D^2 \rho^2 R^2}{4 n^2} .
\end{equation}
In obtaining this equation of motion we have squared some quantities, so there might be spurious solutions, that is, there could be solutions of (\ref{eqnmotion1}) which are not solutions of (\ref{Israelndima}) for given values of $(\xi_I,\xi_{II})$.
%provided the $\pm$ sign was {\it a priori} specified.
Nevertheless, it can be shown that every solution of (\ref{eqnmotion1}) is a solution of one of the versions of (\ref{Israelndima}), that is, it is also a solution only for {\it certain} value of the pair $(\xi_I,\xi_{II})$.
Therefore, we can find solutions of the junction conditions by solving (\ref{eqnmotion1}) and specifying {\it a posteriori} $(\xi_I,\xi_{II})$ accordingly\footnote{This is a specific class within the general problem of gluing two spherically symmetric spacetimes by means of an hypersurface, as analised in \cite{FST}. See also \cite{GoldwirthKatz} for an analysis on the constraints that energy conditions pose on these constructions.}.
% !!!! don't site FST and do cite Goldwirth Katz!!!!
 %if we were allowed to specify {\it a posteriori} the $\pm$ sign, that is, for a given solution of (\ref{eqnmotion1}).
In this way, for a given solution of the equation of motion $R(\tau)$, the specification that we must made is the following
\begin{eqnarray}
\label{espureoI}
\xi_I = \mbox{sign}(n^2(F_{I}-F_{II})+\kappa_D^2 \rho^2 R^2)=\mbox{sign}(n^2(Q^2_{I}-Q^2_{II})+2n^2(M_{II}-M_I)R^{n-1}+\kappa_D^2 \rho^2 R^{2n}) \\
\label{espureoII}
\xi_{II} = \mbox{sign}(n^2(F_{I}-F_{II})-\kappa_D^2 \rho^2 R^2)=\mbox{sign}(n^2(Q^2_{I}-Q^2_{II})+2n^2(M_{II}-M_I)R^{n-1}-\kappa_D^2 \rho^2 R^{2n}).
%\mbox{sign}(\partial r/\partial \eta|_{\eta=0^+})=\mbox{sign}(n^2(F_{I}-F_{II})-\kappa^2 \rho^2 R^2)=\mbox{sign}(2n^2(M_{II}-M_I)-\kappa^2 \rho^2 R^{n+1}).
\end{eqnarray}
A direct substitution of (\ref{eqnmotion1}) into (\ref{Israelndima}) with these choices for $(\xi_I,\xi_{II})$ proves our point. We stress the fact that (\ref{espureoI}) and (\ref{espureoII}) are valid only if $\rho>0$, otherwise $\mbox{sign}(\rho)$ would appear in the expressions. 
%If we make use of the arbitrariness of the labels $(I,II)$, then we may define the region $I$ as the one such that $F_I>F_{II}$... (consistency of this choice?)
This specification is consistent because the roots of $n^2(F_{I}-F_{II}) \pm \kappa_D^2 \rho^2 R^2$ are always in {\it forbidden regions} ($V(R)>0$) or inside an event horizon $F_{I,II}<0$, as shown in Appendix \ref{specification}. In this way, with these specifications, any solution of the equation of motion (\ref{eqnmotion1}) is a solution of (\ref{Israelndima}). In particular, a collapsing solution of (\ref{eqnmotion1}) is a legitimate solution of the junction conditions. 

Provided $I$ is an interior region, it is worth noticing that solutions where $M_{II}\leq M_I$ that have the standard orientation ($\xi_I=\xi_{II}=+1$) are possible only if $|Q_I|>|Q_{II}|$, which illustrates the fact that they are impossible in the uncharged case or for an uncharged shell. Anyway, if $M_{II}<M_I$ it is not obvious whether they can avoid collapse, because of the fact that there would be a root for both $n^2(F_{I}-F_{II}) \pm \kappa_D^2 \rho^2 R^2$ and these solutions are defined only for $R$ smaller than both of these roots\footnote{As mentioned, these roots must necessarily lie in a forbidden region or inside an event horizon.}, so the radius of a shell with the aforementioned properties is bounded from above. We also remark that if the bulk parameters are the same at both sides, then both bulk regions must be interior, which in turn implies that they must be identical (there would be $Z_2$-symmetry centered at the thin shell).  
%comment on large R limit

\section{Equations of motion and matter models}
\label{sphericalmotion}

In this Section we study general properties of the dynamics of the shell, provided the matter content satisfies the dominant energy condition. In particular, we study the asymptotic dynamics of the shell for both small and large shell radius for shells made of an arbitrary matter model satisfying a couple of reasonable hypothesis besides the energy condition. The analysis of these asymptotics will allow us to infer many qualitative aspects of the general dynamics. We will also perform a more detailed analysis of the general dynamics in the case of linear barotropic fluids.

Let us define the function $\alpha(R)=p(R)/\rho(R)$, which is characteristic of the matter model, then the dominant energy condition implies that $-1\leq \alpha(R) \leq 1$ (while $\rho>0$). We can now write the continuity equation (\ref{conservation}) as
\begin{equation}
(\ln(\rho))'+\frac{n(1+\alpha)}{R}=0,
\end{equation}
 which in turn implies 
\begin{equation}
\label{rho}
\frac{d (\ln(\rho))}{d(\ln(R))}=-n(1+\alpha).
\end{equation}

\subsection{Large $R$ asymptotics}

We now impose that the matter model satisfies the following hypothesis: 
\begin{equation}
\lim_{R\to\infty} \alpha (R) = \alpha_{\infty}.
\end{equation} 

Then, taking into account (\ref{rho}), we have that $\rho$ acquires the asymptotic form  $\rho \approx CR^{-n(1+\alpha_{\infty})}$, where $C$ is a constant, for large enough $R$. The asymptotic behavior of $V(R)$ in the limit $R\to \infty$ can be studied by means of the following scheme.

\begin{table}[h]
\centerline{\includegraphics[width=1.\textwidth]{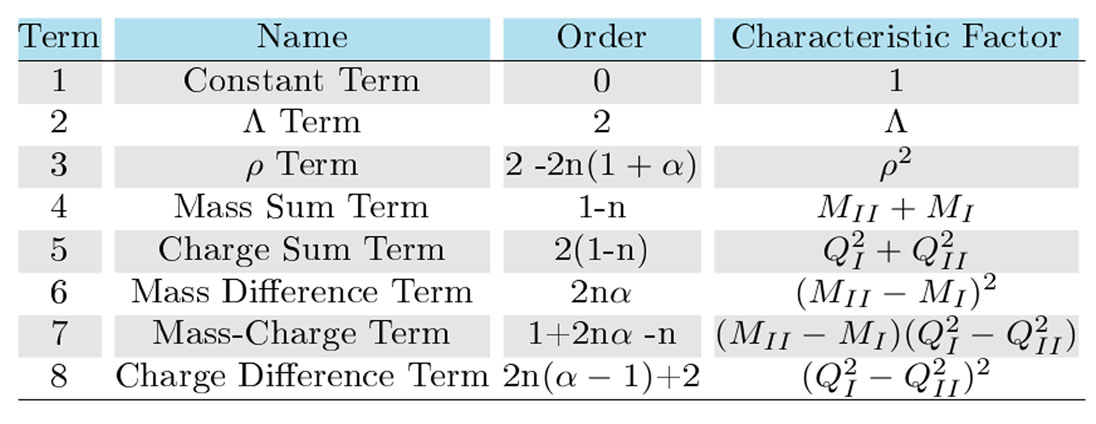}}
\caption{Names assigned to the different terms of the effective potential.\label{terms}}
\end{table}

%CHANGE "EXPRESSION" TO "CHARACTERISTIC FACTOR"

\begin{figure}[h]
\centerline{\includegraphics[width=.5\textwidth]{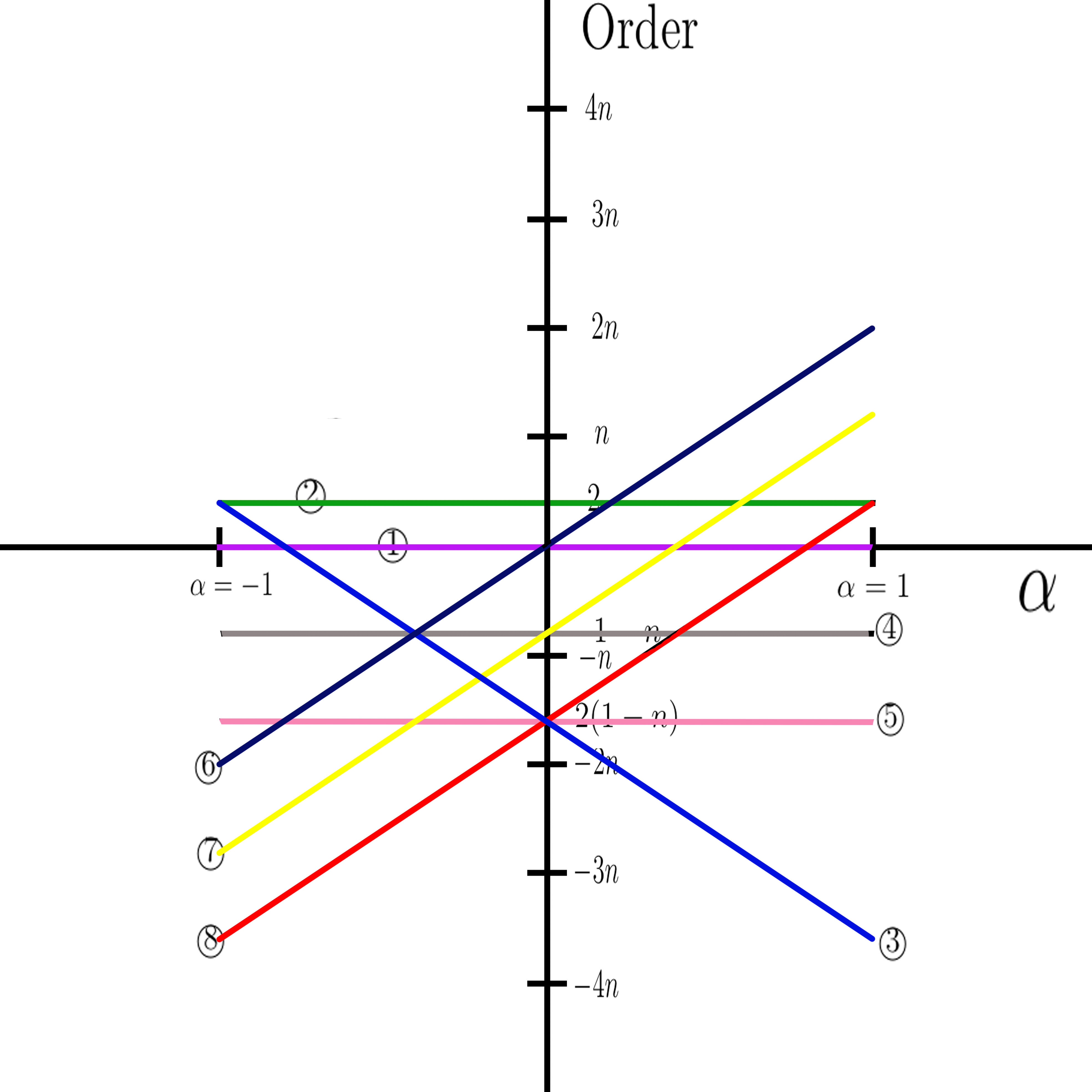}}
\caption{Asymptotic order of the eight terms of the effective potential as functions of $\alpha_{\infty}$ or $\alpha_0$.\label{asymptotics}}

\end{figure}

Table \ref{terms} assigns names and shows the expressions that determine the order of each term of the effective potential, while Figure \ref{asymptotics} illustrates these orders and allows us to determine which term dominates for different values of $\alpha_{\infty}$ in the allowed range. One simply must verify which line(s) is (are) the uppermost one(s) at each point of the horizontal axis. 
We are going to perform this analysis first at full generality (if every term of the potential is non-zero), and then particularise for different cases in which one or more of the terms are canceled out. 

\subsubsection{General case}

In the general case the asymptotics can be described as follows.

\begin{itemize}
\item $\alpha_{\infty} = -1$

In this case both the $\rho$ term and the cosmological constant term dominates at large $R$. 
%We would have $V(R) \approx -\left(\frac{2\Lambda}{n(n+1)} + \frac{\kappa_D^2 C^2}{4 n^2}\right)R^2$. In this way, 
If $\Lambda>-\kappa_D^2 C^2 (n+1)/ (8n)$ (which includes every non-negative value) we would have $V(R) <0$, which implies that {\bf there are unbounded solutions}, for any large enough initial radius. On the other hand, if $\Lambda\leq-\kappa_D^2 C^2 (n+1)/ (8n)$ then $V(R)$ is positive for large $R$ (if the equality holds then the dominant term would be the constant term, which is positive), which implies that {\bf there is a maximum radius} for the shell.

\item $-1<\alpha_{\infty}<1/n$

In this range always dominates the cosmological constant term for large $R$. This means that the possibility of having unbounded solutions is determined by the sign of $\Lambda$ alone: {\bf if it is positive there are unbounded solutions, while if it is negative there is a maximum radius}.

\item $\alpha_{\infty}=1/n$

In this case both the cosmological constant term and the term involving the mass difference between both sides dominate. 
%Then $V(R) \approx -\left(\frac{2\Lambda}{n(n+1)}+ \frac{n^2(M_{II}-M_I)^2}{\kappa_D^2 C^2}\right)R^2$. In this way, 
If $\Lambda>-n^3(n+1)(M_{II}-M_I)^2/(2\kappa_D^2 C^2)$ (which includes every non-negative value), then {\bf there would be unbounded solutions} for any large enough initial radius. On the other hand, if $\Lambda\leq-n^3(n+1)(M_{II}-M_I)^2/(2\kappa_D^2 C^2)$ (including the equality, as in that case the dominant term would be the constant term), then {\bf there is a maximum radius} for the shell.  

\item $1/n < \alpha_{\infty} \leq 1$

In this range the term involving the mass difference dominates. It is always negative, so {\bf there are unbounded solutions}.

\end{itemize}

%DECIDE WHETHER TO PUT IN BOLD BOTH POSSIBILITIES WHEN THERE IS A CRITERIUM

\subsubsection{Asymptotically flat case ($\Lambda=0$)}
\label{simplecase1}

\begin{itemize}

\item $-1\leq\alpha_{\infty}<-(n-1)/n$ 

%Then $V(R) \approx -\frac{\kappa_D^2 \rho^2 R^2}{4n^2} $.
In this range the dominant term is the $\rho$ term, which implies that the solution is {\bf unbounded} provided the initial radius is large enough.

\item $\alpha_{\infty}=-(n-1)/n$

%Then $V(R)\approx 1 - \frac{\kappa_D^2 C^2}{4n^2}$. %
In this case both the $\rho$  term and the constant term are dominant. Then the condition to have unbounded solutions is $\kappa_D C \geq 2n$ (which includes the equality as the subdominant term is the mass sum term, which is negative), otherwise there would be a maximum radius for the shell.

\item $-(n-1)/n<\alpha_{\infty}<0$

%Then $V(R)\approx 1$,
In this range the dominant term is the constant term, which means that {\bf there is a maximum radius for the shell}.

\item $\alpha_{\infty}=0$

%Then $V(R)\approx 1 - \frac{n^2(M_{II}-M_I)^2}{\kappa_D^2 C^2}$. 
In this case both the mass difference term and the constant term dominate, which implies that the condition to have unbounded solutions is $n | M_{II}-M_I |>\kappa_D C$. On the other hand, if $n | M_{II}-M_I |<\kappa_D C$ there is a maximum radius for the shell. If $n | M_{II}-M_I |=\kappa_D C$ then the condition to have unbounded solutions turns out to be $Q_{II}^2 - Q_{I}^2 < M_{II}^2-M_I^2$ (the subdominant terms are the mass sum term and the charge-mass term).

\item $0<\alpha_{\infty}\leq 1$

%Then $V(R)\approx - \frac{n^2(M_{II}-M_I)^2}{\kappa_D^2 \rho^2 R^{2n}} $, 
In this range the dominant term is the mass difference term, which implies that {\bf there is an unbounded solution} for any large enough initial radius.

\end{itemize}

\subsubsection{Equal masses at both sides $M_{II}=M_I$}

\begin{itemize}

\item $\alpha_\infty = -1$

In this case both the cosmological constant term and the $\rho$ term dominate. The condition to have unbounded solutions is $\Lambda>-\kappa_D^2 C^2 (n+1)/ (8n)$ (the equality is not included as the subdominant term is the constant term). 
%(which includes every non-negative value) {\bf there are unbounded solutions}, for any large enough initial radius. On the other hand, if $\Lambda\leq-\kappa_D^2 C^2 (n+1)/ (8n)$ then {\bf there is a maximum radius} for the shell.    

\item $-1<\alpha_\infty < 1$
 
In this range the cosmological constant term dominates, so the possibility of having unbounded solutions is determined by the sign of $\Lambda$ alone. If $\Lambda$ is positive then {\bf there are unbounded solutions}, while in the case of a negative $\Lambda$ {\bf there is a maximum radius for the shell}. 
 
\item $\alpha_\infty = 1$

In this case both the cosmological constant term and the charge difference term dominate. The condition to have unbounded solutions is $\Lambda>-n^3(n+1)(Q_I^2-Q_{II}^2)^2/(8\kappa_D^2C^2)$ (the equality is not included as the subdominant term is the constant term.)
%which includes every non-negative value of $\Lambda$.  On the other hand, if $\Lambda\leq-n^3(n+1)(Q_I^2-Q_{II}^2)^2/(8\kappa_D^2C^2)$ then {\bf there is a maximum radius} for the shell. 

\end{itemize}

\subsubsection{Equal masses at both sides and $\Lambda=0$}

% Either show that it is inconsistent with the Einstein-Maxwell equations, or do it!!!

\begin{itemize}
\item $-1\leq\alpha_{\infty}<-(n-1)/n$ 

In this range the $\rho$ term dominates, so {\bf there are unbounded solutions}.  

\item $\alpha_{\infty}=-(n-1)/n$

In this case both the $\rho$ term and the constant term dominate, so the condition to have unbounded solutions is $\kappa_D C \geq 2n$ (the equality is included as the subdominant term is the mass sum term).

\item  $-(n-1)/n<\alpha_{\infty}<(n-1)/n$

In this range the constant term dominate, so {\bf there is a maximum radius for the shell}.

\item $\alpha_{\infty}=(n-1)/n$ 

In this case both the constant term and the charge difference term dominate, so the condition to have unbounded solutions is $n|Q_I^2-Q_{II}^2|\geq 2\kappa_D C$ (the equality is included as the subdominant term is the mass sum term).

\item $(n-1)/n<\alpha_{\infty}\leq 1$ 

In this range the charge difference term dominates, so {\bf there are unbounded solutions}.

\end{itemize}

%MAY BE WE SHOULD NOT INCLUDE THIS!!

\subsubsection{Equal masses and equal charges at both sides and $\Lambda=0$}

\begin{itemize}

\item $-1\leq\alpha_{\infty}<-(n-1)/n$ 

In this range the $\rho$ term dominates, so {\bf there are unbounded solutions}.  

\item $\alpha_{\infty}=-(n-1)/n$

In this case both the $\rho$ term and the constant term dominate, so the condition to have unbounded solutions is $\kappa_D C \geq 2n$ (the equality is included as the subdominant term is the mass sum term).

\item $-(n-1)/n<\alpha_{\infty}\leq 1$

In this range the constant term dominates, so {\bf there is a maximum radius for the shell}.  

\end{itemize}

The cases in which both charges are equal ($Q_I=Q_{II}$, including the uncharged case) but not both masses ($M_I\neq M_{II}$) are not specifically considered here because they are already included in the ``general case'' and in the ``asymptotically flat case'', as these analysis are completely independent from the charge terms. The charges make no difference in the qualitative aspects of the large $R$ asymptotic limit unless the masses at both sides are equal.

\subsection{Small $R$ asymptotics}
\label{smallr}

Analogously, we are going to consider the small $R$ asymptotics in order to determine whether the shell can collapse to zero size, or, on the contrary, there is a non-zero minimum radius for the shell, which would imply a rebound if the shell reaches this radius. In this way, we impose the following condition on the matter content of the shell
\begin{equation}
\lim_{R\to 0} \alpha(R) = \alpha_0 .
\end{equation}
This implies that the matter energy density goes like $\rho \approx C R^{-n(1+\alpha_0)}$ for $R$ small enough. In order to determine which terms dominate $V(R)$ in this limit we can also use figure \ref{asymptotics}. In this case we must look at which line(s) is(are) the {\bf lowermost} at each point of the horizontal axis.

\subsubsection{General case}

\begin{itemize}

\item $-1\leq \alpha_0 < 0$

In this range the dominant term %in $V(R)$
is the charge difference term, which is negative. In this way, there is a {\bf collapsing solution} for any small enough initial radius.

\item $\alpha_0 =0$ 

In this case the dominant terms are the charge difference term, the charge sum term, and the $\rho$ term. Therefore, the criteria that determines the possibility of having collapse solutions is given by $(Q_I^2+Q_{II}^2)/2-n^2(Q_I^2-Q_{II}^2)^2/(4\kappa_D^2C^2)-\kappa_D^2 C^2/(4n^2)<0$. In case that $(Q_I^2+Q_{II}^2)/2-n^2(Q_I^2-Q_{II}^2)^2/(4\kappa_D^2C^2)-\kappa_D^2 C^2/(4n^2)=0$, then the condition to have collapsing solutions turns out to be $Q_{II}^2-Q_I^2<\frac{(M_I+M_{II})\kappa_D^2C^2}{ (M_{II}-M_I)n^2}$.

\item $0<\alpha_0\leq 1$

In this range the $\rho$ term dominates, which is always negative, so {\bf there are collapsing solutions}.

\end{itemize}

\subsubsection{Equal charges at both sides ($Q_{II}=Q_I$)}

\begin{itemize}

\item $-1\leq \alpha_0 < - (n-1)/n$

In this range the dominant term is the mass difference term, which is negative, so {\bf there are always collapse solutions} for a small enough initial radius.

\item $\alpha_0=-(n-1)/n$

In this case the dominant terms are both the mass difference term and the charge term. In this way, the condition to have collapsing solutions is $Q^2\leq n^2 (M_{II}-M_I)^2/\kappa_D^2 C^2$ (if the equality holds, then the dominant term is the mass sum term, which is always negative). Otherwise, there is a minimum radius for the shell. 

\item $-(n-1)/n<\alpha_0 < 0$

In this range the charge term is the dominant term, which is positive. This implies that a collapse is not possible, and {\bf there is a minimum radius for the shell}. 

\item $\alpha_0=0$

In this case both the charge term and the $\rho$ term are dominant. Then, the condition to have collapsing solutions is $Q^2\leq\kappa_D^2 C^2/4n^2$ (like in the previous cases, the subdominant term is the mass sum term, which is negative). Otherwise, there is a minimum radius for the shell. 

\item $0<\alpha_0 \leq 1$

In this range the dominant term is the $\rho$ term, which is negative. So, there is a {\bf collapse solution} for any small enough initial radius.

\end{itemize}

\subsubsection{Equal charges and equal masses at both sides ($Q_{II}=Q_I$ and $M_{II}=M_I$)}

\begin{itemize}
\item $-1\leq\alpha_0 < 0$

In this range the dominant term is the charge term, so {\bf there is a minimum radius for the shell}. 

\item $\alpha_0 =0$

In this case the dominant terms are the charge term and the $\rho$ term. Then, the condition to have collapsing solutions is $Q^2\leq\kappa_D^2 C^2/4n^2$ (the inequality is sharp because the mass sum term is subdominant).

\item $0 < \alpha_0 \leq 1$

In this range the dominant term is the $\rho$ term, so {\bf there are collapsing solutions}.

\end{itemize}

\subsubsection{Uncharged case ($Q_{II}=Q_I=0$)}
\label{simplecase2}

\begin{itemize}
\item $-1\leq \alpha_0 < -(n-1)/2n$

In this range the mass difference term dominates, which implies that there is a {\bf collapse solution} for any small enough initial radius. 

\item $\alpha_0=-(n-1)/2n$ 

In this case three different terms dominate: the mass difference term, the mass sum term and the $\rho$ term. Since all these terms are negative, {\bf there are collapse solutions}. 

\item $-(n-1)/2n<\alpha_0\leq 1$

In this range the $\rho$ term dominates, which also implies that there is a {\bf collapse solution} for any small enough initial radius.  

\end{itemize}

\subsubsection{Equal masses and uncharged case ($Q_{II}=Q_I=0$ and $M_{II}=M_I$)}

\begin{itemize}

\item $-1\leq \alpha_0 < -(n-1)/2n$ 

In this range the mass sum term dominates, which implies that {\bf there are collapse solutions}.

\item $\alpha_0 =-(n-1)/2n$

In this case both the mass sum term and the $\rho$ term dominate. Since both terms are negative, {\bf there also are collapse solutions}.

\item $-(n-1)/2n<\alpha_0\leq 1$

In this range the $\rho$ term dominates, which also implies that there is a {\bf collapse solution} for any small enough initial radius.

\end{itemize}

It is clear that in the uncharged case there always exists the possibility of having a collapse solution. This makes sense from a Newtonian point of view: what may prevent a collapse by creating an infinite potential barrier in the effective potential are the charge terms.  Analogously to the previous subsection, we do not need to consider the case of equal masses and different charges separately, as this case is already included in the ``general case''. An eventual equality of the masses would only play a role in the qualitative aspects of the small $R$ limit only if the charges are also equal.

%Now it is clear that not for every matter model there exist unbounded solutions, so the limit $R \to \infty$ not always makes sense.
%Anyway, this limit, provided it exists, is important because of the fact that for large $R$ we may find Newtonian or low-energy limits for the dynamics of the shell.
%For example in the context of brane-world cosmologies we hope to recover the standard Friedmann equations for $R$ large in some sense. 

We stress the fact that if we have $k$ non-interacting matter fields (so we can write $\rho=\sum_{i=1}^{k}\rho_i$ and $p=\sum_{i=1}^k p_i$), each with its own conservation equation and its corresponding $\alpha_{i \infty}$ and $\alpha_{i0}$, then $\alpha_{\infty}=\min_i\{\alpha_{i \infty}\}$ and $\alpha_0=\max_i\{\alpha_{i0}\}$.
In the next Section we analyse some properties of $V(R)$ for specific matter models.

%These inequalities are also valid for an empty shell with brane tension ($\rho=-p=-b<0$) which do not satisfy dominant energy condition.

%The permitted regions for the function $R(\tau)$ are those in which $V(R)<0$. Those regions depend precisely on the function $\rho(R)$, that is on the matter model and its parameters. We nevertheless can say something about the asymptotics of $V(R)$ according to the general inequalities (\ref{dominant})
%\begin{equation}
%R \rightarrow \infty  \;\;\;\;\;\; V(R) \rightarrow
%\end{equation}
%parece que solo importan los �ltimos dos t�rminos, para el infinito. Mentira, el 1/2 puede importar tambi�n (E-V)

% y para r tendiendo a cero tambi�n importar�an s�lo estos t�rminos..

\section{\textbf{Barotropic fluid with equation of state $p=\omega\rho$}}
\label{barotropic}

For this family of matter models, equation (\ref{conservation}) implies 
\begin{equation}
\rho(R)= \rho_0 \left(\frac{R_0}{R}\right)^{n(1+\omega)}.
\end{equation}
And in this case $V(R)$ can be written

\begin{equation}
\label{potential2}
V(R)= 1 -\frac{2\Lambda R^2}{n(n+1)}- \frac{M_I + M_{II}}{R^{n-1}} + \frac{Q_I^2+Q_{II}^2}{2R^{2(n-1)}} - \left(\frac{n(M_{II}-M_{I})}{\kappa_D C R^{-n\omega}}+\frac{n(Q_I^2-Q_{II}^2)}{2\kappa_D CR^{n-n\omega-1}}\right)^2-\frac{\kappa_D^2 C^2 R^{2-2n(1+\omega)}}{4 n^2} ,
\end{equation}
where $C\equiv \rho_0 R_0^{n(1+\omega)}$. We have $\alpha(R)=\alpha_{\infty}=\alpha_0=\omega$ and the dominant energy condition implies $-1\leq \omega \leq 1$.

The aim of this Section is to describe the general features of the dynamics of the shell that can be obtained from the asymptotic behaviour we already analysed. This Section is, in a sense, a combination of both asymptotic analysis that can be made because we defined an equation of state for the matter-energy content of the shell. By knowing the dominant terms at both extremes of our domain in $R$ (the positive real numbers) we can obtain sufficient conditions to decide whether $V(R)$ has a root or a local extremum. We can also get necessary conditions for monotonicity of the potential; then, if those conditions are met, we analyse the first derivative of $V(R)$.  

We stress the fact that there are three values of $\omega$ of particular importance because of their significance in cosmology and astrophysics: $\omega=-1$ represents a cosmological constant fluid or a {\it surface tension}, $\omega=0$ represents dust, and $\omega=1/n$ represents a photon gas.

%are you sure that monotonicity is impossible???are you sure that there might not be forbidden regions?
\subsubsection{General case}
%COMPARE WITH EIROA-SIMEONE 13/12

\begin{itemize}
\item $\omega = -1$

%In this case the potential will not be monotonic. In this way,
In this case if $\Lambda>-\kappa_D^2C^2(n+1)/(8n)$ there would be at least one local maximum for $V(R)$ and there is a subset of the parameter space in which there are no forbidden regions (that is, $V(R)<0$ in its entire range). On the other hand, if $\Lambda\leq-\kappa_D^2C^2(n+1)/(8n)$ there would be a maximum radius and at least one root for $V(R)$.

\item $-1< \omega <0$

%In this range the potential will not be monotonic. In this way,
In this range if $\Lambda>0$ there would be at least one local maximum for $V(R)$ and there is subset of the parameter space in which there are no forbidden regions for the shell. On the other hand, if $\Lambda<0$ there would be a maximum radius and at least one root for $V(R)$.

\item $\omega =0$

%REVISE THIS CASE WITH DANIEL!!!!!! 13/12
In this case if $\Lambda>0$ and $(Q_I^2+Q_{II}^2)/2 - n^2(Q_I^2-Q_{II}^2)^2/(4\kappa^2_D C^2)-\kappa_D^2 C^2/(4 n^2)>0$ there would be a minimum radius for the shell and at least one root for $V(R)$; if $\Lambda>0$ and $(Q_I^2+Q_{II}^2)/2 - n^2(Q_I^2-Q_{II}^2)^2/(4\kappa^2_D C^2)-\kappa_D^2 C^2/(4 n^2)<0$ 
%the potential will not be monotonic and 
there would be at least one local maximum and there is a subset of the parameter space in which there are no forbidden regions for the shell: for example, this is the case if $|M_{II}-M_{I}|>\kappa_D C/n$ and $M_{I}+M_{II}>n^2(M_{II}-M_{I})(Q^2_{II}-Q^2_{I})/(\kappa^2_D C^2)$. On the other hand, if $\Lambda<0$ and $(Q_I^2+Q_{II}^2)/2 - n^2(Q_I^2-Q_{II}^2)^2/(4\kappa^2_D C^2)-\kappa_D^2 C^2/(4 n^2)>0$ %the potential will not be monotonous and 
there would be at least one local minimum and all possible solutions are oscillating: for example, this is the case if $|M_{II}-M_{I}|>\kappa_D C/n$ and $|\Lambda|$ is sufficiently small, while for certain subset of the parameter space there is no solution at all. 
%and $(M_{II}-M_{I})(Q^2_{I}-Q^2_{II})<0$. 
Finally, if $\Lambda<0$ and $(Q_I^2+Q_{II}^2)/2 - n^2(Q_I^2-Q_{II}^2)^2/(4\kappa^2_D C^2)-\kappa_D^2 C^2/(4 n^2)<0$ 
there would be a maximum radius and at least one root;
%and $Q_I^2<Q_{II}^2$ the potential will not be monotonous, we will have a maximum radius  and there is at least one root; if $\Lambda<0$ and $(Q_I^2+Q_{II}^2)/2 - n^2(Q_I^2-Q_{II}^2)^2/(4\kappa^2_D C^2)-\kappa_D^2 C^2/(4 n^2)<0$ and 
moreover, if at the same time $Q_I^2>Q_{II}^2$ then the potential would be {\bf monotonically increasing}, with a single root (which is the maximum radius), so the final outcome of the dynamics would always be a {\bf collapse}.

\item $0<\omega <1/n$

%In this range the potential will not be monotonous. In this way,
In this range if $\Lambda>0$ there would be at least one local maximum for $V(R)$ and there is a subset of the parameter space in which there are no forbidden regions for the shell. On the other hand, if $\Lambda<0$ there would be a maximum radius and at least one root.

\item $\omega =1/n$

%In this case the potential will not be monotonous. In this way, 
In this case if $\Lambda>-n^3(n+1)(M_{II}-M_I)^2/(2\kappa^2_DC^2)$ there would be at least one local maximum for the potential and there is a subset of the parameter space in which there are no forbidden regions for the shell. On the other hand, if $\Lambda\leq-n^3(n+1)(M_{II}-M_I)^2/(2\kappa^2_DC^2)$ there would be a maximum radius and at least one root.

\item $1/n< \omega \leq 1$

In this range there would be at least one local maximum for $V(R)$ and there is a subset of the parameter space in which there are no forbidden regions for the shell.
\end{itemize}

\subsubsection{Asymptotically flat case ($\Lambda=0$)}

\begin{itemize}

\item $-1\leq \omega <-(n-1)/n$

In this range there would be at least one local maximum for $V(R)$ and there is a subset of the parameter space in which there are no forbidden regions.

\item $\omega =-(n-1)/n$

In this case if $\kappa_DC \geq 2n$ there is a subset of the parameter space in which there are no forbidden regions. Otherwise, there is at least one root and a maximum radius for the shell.

\item $-(n-1)/n <\omega <0$

In this range there is at least one root for $V(R)$ and a maximum radius for the shell.

\item $\omega =0$

This is the general setting considered in \cite{GaoLemos}. It also includes example 3 of Section 4 of \cite{EiroaSimeone} (a charged dust bubble). It has the particular advantage that the potential takes the form of a quadratic equation in $x=R^{-(n-1)}$ ($V(x)=cx^2+bx+a$) while being physically relevant. The possible dynamics for potentials of this form is explained in appendix \ref{asymptotics}. For this case we have

\begin{equation*}
\begin{aligned}[b]
c=\frac{Q_I^2+Q_{II}^2}{2} -\frac{n^2(Q_I^2-Q_{II}^2)^2}{4\kappa^2_D C^2}-\frac{\kappa_D^2 C^2 }{4 n^2},
\end{aligned}
\qquad
\begin{gathered}[t]
a=1- \frac{n^2(M_{II}-M_{I})^2}{\kappa^2_D C^2}\\
\end{gathered}
\end{equation*}

%$$
%c=\frac{Q_I^2+Q_{II}^2}{2} -\frac{n^2(Q_I^2-Q_{II}^2)^2}{4\kappa^2_D C^2}-\frac{\kappa_D^2 C^2 }{4 n^2} 
%$$

$$
b=- (M_I + M_{II}) -\frac{n^2(M_{II}-M_{I})(Q_I^2-Q_{II}^2)}{\kappa^2_D C^2}
$$

%$$
%a=1- \frac{n^2(M_{II}-M_{I})^2}{\kappa^2_D C^2}
%$$

$$
\Delta=4M_I	M_{II}- \frac{2n^2(Q_I^2M_I+Q_{II}^2M_{II})(M_I+M_{II})}{\kappa_D^2C^2}-2(Q_I^2+Q^2_{II})+\frac{n^2(Q_I^2-Q_{II})^2}{\kappa_D^2C^2}+ \frac{\kappa_D^2C^2}{n^2}
$$.

In this relatively general context each coefficient can acquire any sign, so all the subcases of the appendix \ref{quadratic} are possible. Of particular interest are the situations where the motion is oscillatory (subcase E in Appendix \ref{quadratic}), this is analysed in Section 4 of \cite{GaoLemos}. As explained there, if there is an interior black hole and the shell has the standard orientation then these oscillations actually represent a collapse: there can only be oscillating solutions if the exterior solution also corresponds to a black hole and in that case the shell must enter the event horizon and reemerge in another asymptotically flat region of the extended exterior Reissner-Nordstr\"om spacetime. This will be further explained in Section \ref{final}.

\item $0< \omega \leq 1$

In this range the potential would have a local maximum for $V(R)$ and for certain subset of the parameter space there would not be any forbidden region.
\end{itemize}

\subsubsection{Equal charges at both sides ($Q_{II}=Q_I$)}

\begin{itemize}

\item $\omega = -1$

In this case if $\Lambda>-\kappa_D^2C^2(n+1)/(8n)$ there would be a local maximum for the potential and for certain subset of the parameter space there would not be any forbidden region. On the other hand, if $\Lambda<-\kappa_D^2C^2(n+1)/(8n)$  there would be at least one root and a maximum radius for the shell.

\item $-1< \omega <-(n-1)/n$

In this range if $\Lambda>0$ there would be a local maximum for the potential and for certain subset of the parameter space there would not be any forbidden region. On the other hand, if $\Lambda<0$ there would be at least one root and a maximum radius for the shell.

\item $\omega =-(n-1)/n$

In this case if $\Lambda>0$ and $|Q|>n|M_{II}-M_{I}|/\kappa_D C$ there would be at least one root and a minimum radius for the shell; if $\Lambda>0$ and $|Q|<n|M_{II}-M_{I}|/\kappa_D C$ the potential would have a local maximum and there is a subset of the parameter space in which there are no forbidden regions: for example if $4n^2<\kappa^2_D C^2$. On the other hand, if $\Lambda<0$ and $|Q|>n|M_{II}-M_{I}|/\kappa_D C$ then the potential would have a local minimum, depending on the parameters there are either oscillating solutions or no solutions at all; if $\Lambda<0$ and $|Q|<n|M_{II}-M_{I}|/\kappa_D C$ then the potential would be {\bf monotonically increasing}, there would be a single root, at the maximum radius, so the final outcome of the evolution would always be a {\bf collapse}.

\item $-(n-1)/n <\omega <0$

In this range if $\Lambda>0$ there would be at least one root and a minimum radius. On the other hand if $\Lambda<0$ there would be a local minimum for the potential,depending on the parameters there are either oscillating solutions or no solutions at all.

\item $\omega =0$

In this case if $\Lambda>0$ and $|Q|>\kappa_D C /(2 n)$ there would be at least one root and a minimum radius; if $\Lambda>0$ and $|Q|<\kappa_D C /(2 n)$ the potential would have a local maximum and there is a subset of the parameter space in which there are no forbidden regions: for example if $|M_{II}-M_{I}|>\kappa_D C/n$. On the other hand, if $\Lambda<0$ and $|Q|>\kappa_D C /(2 n)$ there would be a local minimum for the potential, depending on the parameters there are either oscillating solutions or no solutions at all; if $\Lambda<0$ and $|Q|<\kappa_D C /(2 n)$ then the potential is {\bf monotonically increasing}, there is a single root, at the maximum radius, so the outcome of the evolution would always be a {\bf collapse}.

\item $0< \omega <1/n$

In this range if $\Lambda>0$ there is a local maximum for the potential and for certain subset of the parameter space there would not be any forbidden regions. On the other hand, if $\Lambda<0$ there would be at least one root and a maximum radius.

\item $\omega =1/n$

In this case if $\Lambda>-n^3(n+1)(M_{II}-M_I)^2/2\kappa^2_DC^2$ there would be a local maximum for the potential and for certain subset of the parameter space there would not be any forbidden regions. On the other hand, if $\Lambda<-n^3(n+1)(M_{II}-M_I)^2/2\kappa^2_DC^2$ there would be at least one root and a maximum radius.

\item $1/n< \omega \leq 1$

In this range there would be a local maximum for the potential and for certain subset of the parameter space there would not be any forbidden regions.
\end{itemize}

\subsubsection{Equal charges at both sides ($Q_{II}=Q_I$) and $\Lambda=0$ }

\begin{itemize}

\item $-1\leq \omega <-(n-1)/n$

In this range there would be a local maximum for the potential and for certain subset of the parameter space there would not be any forbidden regions.

\item $ \omega =-(n-1)/n$

For this case, the potential takes the form of a quadratic equation in $x=R^{-(n-1)}$.

\begin{equation*}
\begin{aligned}[b]
c=Q^2 - \frac{n^2(M_{II}-M_{I})^2}{\kappa^2_D C^2},
\end{aligned}
\qquad
\begin{gathered}[t]
b=-(M_I + M_{II}),\\
\end{gathered}
\qquad
\begin{gathered}[t]
a=1-\frac{\kappa_D^2 C^2}{4 n^2 }\\
\end{gathered}
\end{equation*}
$$
\Delta=4M_IM_{II}-4Q^2+\frac{4n^2(M_{II}-M_I)^2}{\kappa_D^2C^2}+\frac{Q^2\kappa_D^2C^2}{n^2}
$$

where if $c>0$: 
\begin{enumerate}
\item[$\circ$]$\Delta>0$ and $a\leq 0$ we have the subcase C of the appendix \ref{quadratic}. On the other hand, if $a> 0$ we have the subcase E.
\item[$\circ$] $\Delta<0$, we have the subcase D of the appendix \ref{quadratic}.
\item[$\circ$] $\Delta=0$, we have the subcase E* of the appendix \ref{quadratic}.
\end{enumerate}

If $c<0$: 
\begin{enumerate}
\item[$\circ$]$\Delta>0$ and $a\leq 0$ we have the subcase A of the appendix \ref{quadratic}. On the other hand, if $a >0$ we have the subcase B.
\end{enumerate}

%In this case if $Q^2>n|M_{II}-M_I|/\kappa_DC$ and $2n<\kappa_DC$ there is a single root, which is a maximum radius. On the other hand, if $Q^2<n|M_{II}-M_I|/\kappa_DC$ and $2n>\kappa_DC$ then the potential is {\bf monotonically increasing}, there is a single root, which is a maximum radius, so the outcome of the evolution would always be a {\bf collapse}; if $Q^2<n|M_{II}-M_I|/\kappa_DC$ and $2n<\kappa_DC$ then the potential is {\bf monotonically increasing} and the shell can either expand indefinitely or collapse.

\item $ -(n-1)/n <\omega <0 $

In this range for certain subset of the parameter space there would not be any forbidden regions.

\item $\omega =0 $

This case is example 1 of Section 4 of \cite{EiroaSimeone}. For this case, the potential takes the form of a quadratic equation in $x=R^{-(n-1)}$.

\begin{equation*}
\begin{aligned}[b]
c=Q^2 - \frac{\kappa^2_D C^2}{4n^2},
\end{aligned}
\qquad
\begin{gathered}[t]
b=-(M_I + M_{II}),\\
\end{gathered}
\qquad
\begin{gathered}[t]
a=1- \frac{n^2(M_{II}-M_{I})^2}{\kappa^2_D C^2}\\
\end{gathered}
\end{equation*}

$$
\Delta=4M_IM_{II}-4Q^2+\frac{\kappa^2_D C^2}{n^2}+\frac{4Q^2n^2(M_{II}-M_I)^2}{\kappa_D^2C^2}
$$

where if $c>0$: 
\begin{enumerate}
\item[$\circ$]$\Delta>0$ and $a\leq 0$ we have the subcase C of the appendix \ref{quadratic}. On the other hand, if $a> 0$ we have the subcase E.
\item[$\circ$] $\Delta<0$, we have the subcase D of the appendix \ref{quadratic}.
\item[$\circ$] $\Delta=0$, we have the subcase E* of the appendix \ref{quadratic}.
\end{enumerate}

If $c<0$:
\begin{enumerate}
\item [$\circ$]$\Delta>0$ and $a\leq 0$ we have the subcase A of the appendix \ref{quadratic}. On the other hand, if $a >0$ we have the subcase B.
\end{enumerate}

%In this case if $Q>\kappa_DC/(2n)$ and $|M_{II}-M_I|>\kappa_DC/n$ there is a single root, which is a minimum radius. If $Q<\kappa_DC/(2n)$ then the potential would be {\bf monotonically increasing}. If at the same time  $|M_{II}-M_I|<\kappa_DC/n$ then there would be a single root for $V(R)$, which is a maximum radius, and the final outcome of the evolution would always be a {\bf collapse}; on the other hand, if $|M_{II}-M_I|>\kappa_DC/n$ then the shell can either expand indefinitely or collapse and there would not be any forbidden region for the shell.

\item $ 0 <\omega \leq 1 $

In this range there would be a local maximum for the potential and for certain subset of the parameter space there would not be any forbidden regions.

\end{itemize}

\subsubsection{Uncharged case ($Q_{II}=Q_I=0$)}

\begin{itemize}

\item $ \omega =-1 $

In this case if $\Lambda>(n+1)\kappa^2_DC^2/(8n)$ there would be at least one local maximum for $V(R)$, there is a subset of the parameter space in which there are no forbidden regions for the shell. On the other hand if $\Lambda<(n+1)\kappa^2_DC^2/(8n)$  then the potential would be {\bf monotonically increasing}, with a single root (which is the maximum radius), so the final outcome of the dynamics would always be a {\bf collapse}.

\item $-1< \omega <-(n-1)/n $

In this range if $\Lambda>0$ there would be at least one local maximum for $V(R)$, there is a subset of the parameter space in which there are no forbidden regions for the shell. On the other hand, if $\Lambda<0$ there would be a maximum radius and at least one root.

\item $-(n-1)/n\leq \omega \leq 0 $

In this range if $\Lambda>0$ there would be at least one local maximum for $V(R)$ and there is a subset of the parameter space in which there are no forbidden regions for the shell. On the other hand, if $\Lambda<0$ then the potential would be {\bf monotonically increasing}, there would be a single root, at the maximum radius, so the final outcome of the evolution would always be a {\bf collapse}. This item includes example 4 of Section 4 of \cite{EiroaSimeone} (a dust shell in a cosmological constant background).

\item $0< \omega <1/n $

In this range if $\Lambda>0$ there would be at least one local maximum for $V(R)$ and there is a subset of the parameter space in which there are no forbidden regions for the shell. On the other hand, if $\Lambda<0$ there would be a maximum radius and at least one root.

\item $\omega =1/n $

In this case if $\Lambda>n^3(n+1)(M_{II}-M_I)^2/2\kappa^2_DC^2$ there would be at least one local maximum for $V(R)$ and there is a subset of the parameter space in which there are no forbidden regions for the shell. On the other hand, if $\Lambda<n^3(n+1)(M_{II}-M_I)^2/2\kappa^2_DC^2$ then the potential would be {\bf monotonically increasing}, with a single root (which is the maximum radius), so the final outcome of the dynamics would always be a {\bf collapse}.

\item $1/n <\omega \leq 1 $

In this range there would be a local maximum for the potential and for certain subset of the parameter space there would not be any forbidden regions.

\end{itemize}

\subsubsection{Uncharged case ($Q_{II}=Q_I=0$) with $\Lambda=0$}

This is example 2 of Section 4 of \cite{EiroaSimeone}.

\begin{itemize}

\item $-1\leq \omega <-(n-1)/n $

In this range there would be a local maximum for the potential and there is a subset of the parameter space in which there are no forbidden regions for the shell.

\item $\omega =-(n-1)/n $

For this case, the potential takes the form of a quadratic equation in $x=R^{-(n-1)}$.

\begin{equation*}
\begin{aligned}[b]
c=- \frac{n^2(M_{II}-M_{I})^2}{\kappa^2_D C^2},
\end{aligned}
\qquad
\begin{gathered}[t]
b=-(M_I + M_{II}),\\
\end{gathered}
\qquad
\begin{gathered}[t]
a=1- \frac{\kappa^2_D C^2}{4n^2}\\
\end{gathered}
\end{equation*}

$$
\Delta=4M_IM_{II}+\frac{4n^2(M_{II}-M_I)^2}{\kappa_D^2C^2}
$$

In this case:
\begin{enumerate}
\item[$\circ$] if $a\leq 0$ we have the subcase A of the appendix \ref{quadratic}. On the other hand, if $a >0$ we have the subcase B.
\end{enumerate}

%In this case the potential would be {\bf monotonically increasing}. If $2n<\kappa_DC$ then there would be a single root for $V(R)$, which is a maximum radius, and the final outcome of the evolution would always be a {\bf collapse}. On the other hand, if $2n>\kappa_DC$ then the shell can either expand indefinitely or collapse and there would not be any forbidden region for the shell.

\item $-(n-1)/n< \omega <0$

In this range the potential would be {\bf monotonically increasing}, with a single root (which is the maximum radius), so the final outcome of the dynamics would always be a {\bf collapse}.

\item $\omega =0$

For this case, the potential takes the form of a quadratic equation in $x=R^{-(n-1)}$.

\begin{equation*}
\begin{aligned}[b]
c=- \frac{\kappa^2_D C^2}{4n^2},
\end{aligned}
\qquad
\begin{gathered}[t]
b=-(M_I + M_{II}),\\
\end{gathered}
\qquad
\begin{gathered}[t]
a=1- \frac{n^2(M_{II}-M_{I})^2}{\kappa^2_D C^2}\\
\end{gathered}
\end{equation*}

$$
\Delta=4M_IM_{II}+\frac{\kappa_D^2C^2}{n^2}
$$

In this case:
\begin{enumerate}
\item[$\circ$] if $a\leq 0$ we have the subcase A of the appendix \ref{quadratic}. On the other hand, if $a >0$ we have the subcase B.
\end{enumerate}

\item $0 <\omega \leq 1 $

In this range there would be a local maximum for the potential and there is a subset of the parameter space in which there are no forbidden regions for the shell
\end{itemize}

\subsubsection{Equal masses at both sides $M_{II}=M_I$}

\begin{itemize}
\item $\omega = -1$

In this range if $\Lambda>-\kappa_D^2C^2(n+1)/(8n)$ there would be at least one local maximum for $V(R)$ and there is a subset of the parameter space in which there are no forbidden regions for the shell. On the other hand, if $\Lambda<-\kappa_D^2C^2(n+1)/(8n)$ there would be a maximum radius and at least one root.

\item $-1< \omega <0$

In this range if $\Lambda>0$ there would be at least one local maximum for $V(R)$ and there is a subset of the parameter space in which there are no forbidden regions for the shell. On the other hand, if $\Lambda<0$ there would be a maximum radius and at least one root.

\item $\omega =0$

In this case if $\Lambda>0$ and $(Q_I^2+Q_{II}^2)/2 - n^2(Q_I^2-Q_{II}^2)^2/(4\kappa^2_D C^2)-\kappa_D^2 C^2/(4 n^2)>0$ there would be a minimum radius for the shell and at least one root for $V(R)$; if $\Lambda>0$ and $(Q_I^2+Q_{II}^2)/2 - n^2(Q_I^2-Q_{II}^2)^2/(4\kappa^2_D C^2)-\kappa_D^2 C^2/(4 n^2)<0$ 
%the potential will not be monotonic and 
there would be at least one local maximum and there is a subset of the parameter space in which there are no forbidden regions for the shell. On the other hand, if $\Lambda<0$ and $(Q_I^2+Q_{II}^2)/2 - n^2(Q_I^2-Q_{II}^2)^2/(4\kappa^2_D C^2)-\kappa_D^2 C^2/(4 n^2)>0$ %the potential will not be monotonous and 
there would be at least one local minimum, depending on the parameters there are either oscillating solutions or no solutions at all. Finally, if $\Lambda<0$ and $(Q_I^2+Q_{II}^2)/2 - n^2(Q_I^2-Q_{II}^2)^2/(4\kappa^2_D C^2)-\kappa_D^2 C^2/(4 n^2)<0$ 
there would be a maximum radius and at least one root;
%and $Q_I^2<Q_{II}^2$ the potential will not be monotonous, we will have a maximum radius  and there is at least one root; if $\Lambda<0$ and $(Q_I^2+Q_{II}^2)/2 - n^2(Q_I^2-Q_{II}^2)^2/(4\kappa^2_D C^2)-\kappa_D^2 C^2/(4 n^2)<0$ and 
moreover, if at the same time $Q_I^2>Q_{II}^2$ then the potential would be {\bf monotonically increasing}, with a single root (which is the maximum radius), so the final outcome of the dynamics would always be a {\bf collapse}.

\item $0<\omega <1$

In this range if $\Lambda>0$ there would be at least one local maximum for $V(R)$ and there is subset of the parameter space in which there are no forbidden regions for the shell. On the other hand, if $\Lambda<0$ there would be a maximum radius and at least one root.

\item $\omega =1$

In this case if $\Lambda>-n^3(n+1)(Q_I^2-Q^2_{II})^2/(8\kappa^2_DC^2)$ there would be at least one local maximum for $V(R)$ and there is a subset of the parameter space in which there are no forbidden regions for the shell. On the other hand, if $\Lambda<-n^3(n+1)(Q_I^2-Q^2_{II})^2/(8\kappa^2_DC^2)$ there would be a maximum radius and at least one root.
 
\end{itemize}

\subsubsection{Equal masses at both sides and $\Lambda=0$}

\begin{itemize}

\item $-1\leq \omega <-(n-1)/n$

In this range there would be a local maximum for the potential and for certain subset of the parameter space there would not be any forbidden regions.

\item $\omega =-(n-1)/n$

In this case if $\kappa_DC \geq 2n$ there is a subset of the parameter space in which there are no forbidden regions. On the other hand, if $\kappa_DC < 2n$ there is at least one root for $V(R)$ and a maximum radius for the shell.

\item $-(n-1)/n <\omega <0$

In this range there is at least one root for $V(R)$ and a maximum radius for the shell.

\item $\omega =0$

For this case, the potential takes the form of a quadratic equation in $x=R^{-(n-1)}$.

\begin{equation*}
\begin{aligned}[b]
c=\frac{Q_I^2+Q_{II}^2}{2} -\frac{n^2(Q_I^2-Q_{II}^2)^2}{4\kappa^2_D C^2}-\frac{\kappa_D^2 C^2}{4 n^2 },
\end{aligned}
\qquad
\begin{gathered}[t]
b=-2M,\\
\end{gathered}
\qquad
\begin{gathered}[t]
a=1\\
\end{gathered}
\end{equation*}

$$
\Delta=4M^2-2(Q^2_I+Q_{II}^2)+\frac{n^2(Q_I^2-Q_{II}^2)^2}{\kappa_D^2C^2}+\frac{\kappa_D^2C^2}{n^2}
$$

where if $c>0$: 
\begin{enumerate}
\item[$\circ$]$\Delta>0$,  we have the subcase E of the appendix \ref{quadratic}.
\item[$\circ$] $\Delta<0$, we have the subcase D of the appendix \ref{quadratic}.
\item[$\circ$] $\Delta=0$, we have the subcase E* of the appendix \ref{quadratic}.
\end{enumerate}

If $c<0$: 
\begin{enumerate}
\item[$\circ$] we have the subcase B of the appendix \ref{quadratic}.
\end{enumerate}

%In this case if $(Q_I^2+Q_{II})^2/2- n^2(Q_I^2-Q_{II}^2)^2/4\kappa^2_D C^2-\kappa_D^2 C^2/(4n^2)>0$ there would be a minimum radius for the shell and at least one root for $V(R)$; if $(Q_I^2+Q_{II})^2/2- n^2(Q_I^2-Q_{II}^2)^2/4\kappa^2_D C^2-\kappa_D^2 C^2/(4n^2)<0$ and $M_I + M_{II} + n^2(M_{II}-M_{I})(Q_I^2-Q_{II}^2)/\kappa^2_D C^2>0$ then the potential would be {\bf monotonically increasing}, with a single root (which is the maximum radius), so the final outcome of the dynamics would always be a {\bf collapse}.
 
\item $0< \omega <(n-1)/n $

In this range there is at least one root for $V(R)$ and a maximum radius for the shell.

\item $\omega=(n-1)/n $

In this case if $|Q_I^2-Q_{II}^2|<2\kappa_D C/n$ there is at least one root for $V(R)$ and a maximum radius for the shell. If $|Q_I^2-Q_{II}^2|\geq2\kappa_D C/n$ there is a subset of the parameter space in which there are no forbidden regions.  

\item $(n-1)/n < \omega  \leq 1$

In this range there would be a local maximum for the potential and for certain subset of the parameter space there would not be any forbidden regions.

\end{itemize}

\subsubsection{Equal masses and charges at both sides}

\begin{itemize}

\item $\omega =-1 $

In this case if $\Lambda>-(n+1)\kappa^2_DC^2/(8n)$ there would be at least one root and a minimum radius. On the other hand if $\Lambda<-(n+1)\kappa^2_DC^2/(8n)$ there would be a local minimum for the potential, depending on the parameters there are either oscillating solutions or no solutions at all.

\item $-1< \omega <0 $

In this range if $\Lambda>0$ there would be at least one root and a minimum radius. On the other hand if $\Lambda<0$ there would be a local minimum for the potential, depending on the parameters there are either oscillating solutions or no solutions at all.

\item $\omega=0$
 
In this case if $\Lambda>0$ and $|Q|>\kappa_D C /(2 n)$ there would be at least one root and a minimum radius; if $\Lambda>0$ and $|Q|<\kappa_D C /(2 n)$ the potential would have a local maximum, there is subset of the parameter space in which there are no forbidden regions for the shell. On the other hand, if $\Lambda<0$ and $|Q|>\kappa_D C /(2 n)$ there would be a local minimum for the potential, depending on the parameters there are either oscillating solutions or no solutions at all; if $\Lambda<0$ and $|Q|<\kappa_D C /(2 n)$ then the potential is {\bf monotonically increasing}, there is a single root, which is a maximum radius, so the outcome of the evolution would always be a {\bf collapse}.

\item $0<\omega \leq 1$

In this range if $\Lambda>0$ there would be at least one local maximum for $V(R)$, there is subset of the parameter space in which there are no forbidden regions for the shell. On the other hand, if $\Lambda<0$ there would be a maximum radius and at least one root.

\end{itemize}

\subsubsection{Equal masses and charges at both sides and $\Lambda=0$}

\begin{itemize}

\item $-1\leq \omega <-(n-1)/n$

In this range there would be a maximum radius and at least one root.

\item $\omega =-(n-1)/n$

For this case, the potential takes the form of a quadratic equation in $x=R^{-(n-1)}$.

\begin{equation*}
\begin{aligned}[b]
c=Q^2,
\end{aligned}
\qquad
\begin{gathered}[t]
b=-2M,\\
\end{gathered}
\qquad
\begin{gathered}[t]
a=1-\frac{\kappa_D^2 C^2}{4 n^2 }\\
\end{gathered}
\end{equation*}

$$
\Delta=4M^2-4Q^2+\frac{Q^2\kappa_D^2C^2}{n^2}
$$

where: 
\begin{enumerate}
\item[$\circ$]$\Delta>0$ and $a\leq 0$ we have the subcase C of the appendix \ref{quadratic}. On the other hand, if $a >0$ we have the subcase E.
\item[$\circ$] $\Delta<0$, we have the subcase D of the appendix \ref{quadratic}.
\item[$\circ$] $\Delta=0$, we have the subcase E* of the appendix \ref{quadratic}.
\end{enumerate}

%\begin{equation}
%V(R)= 1-\frac{\kappa_D^2 C^2}{4 n^2 }  - \frac{2M}{R^{n-1}} + \frac{Q^2}{R^{2(n-1)}}
%\end{equation}

%In this case for certain subset of the parameter space there would not be any forbidden regions. If $2n<K_DC$  there would be a maximum radius and at least one root.

\item $-(n-1)/n<\omega <0$

In this range there is a local minimum for the potential. Depending on the parameters there are either no solutions, oscillating solutions or a static solution.

\item $\omega =0$

For this case, the potential takes the form of a quadratic equation in $x=R^{-(n-1)}$.

\begin{equation*}
\begin{aligned}[b]
c=Q^2-\frac{\kappa_D^2 C^2}{4 n^2},
\end{aligned}
\qquad
\begin{gathered}[t]
b=-2M,\\
\end{gathered}
\qquad
\begin{gathered}[t]
a=1\\
\end{gathered}
\end{equation*}

$$
\Delta=4M^2-4Q^2+\frac{\kappa_D^2C^2}{n^2}
$$

where if $c>0$: 
\begin{enumerate}
\item[$\circ$]$\Delta>0$, we have the subcase E of the appendix \ref{quadratic}.
\item[$\circ$] $\Delta<0$, we have the subcase D of the appendix \ref{quadratic}.
\item[$\circ$] $\Delta=0$, we have the subcase E* of the appendix \ref{quadratic}.
\end{enumerate}

If $c<0$: 
\begin{enumerate}
\item[$\circ$] we have the subcase B of the appendix \ref{quadratic}.
\end{enumerate}

%In this case if $Q>K_DC/(2n)$ for certain subset of the parameter space there would not be any forbidden regions. On the other hand, if $Q<K_DC/(2n)$then the potential would be {\bf monotonically increasing}, with a single root (which is the maximum radius), so the final outcome of the dynamics would always be a {\bf collapse}. 

\item $0<\omega \leq 1$

In this range there is at least one root for $V(R)$ and a maximum radius for the shell.

\end{itemize}

\subsubsection{Equal masses at both sides and uncharged}
 
\begin{itemize}

\item $\omega = -1$

In this case if $\Lambda>-(n+1)\kappa^2_DC^2/(8n)$ there would be at least one local maximum for $V(R)$ and there is subset of the parameter space in which there are no forbidden regions for the shell. On the other hand, if $\Lambda<-(n+1)\kappa^2_DC^2/(8n)$ then the potential would be {\bf monotonically increasing}, with a single root (which is the maximum radius), so the final outcome of the dynamics would always be a {\bf collapse}. 

\item $-1<\omega <-(n-1)/n$

In this range if $\Lambda>0$ there would be at least one local maximum for $V(R)$, there is subset of the parameter space in which there are no forbidden regions for the shell. On the other hand, if $\Lambda<0$ there would be a maximum radius and at least one root.

\item $-(n-1)/n\leq \omega \leq 1$

In this range if $\Lambda>0$ there would be at least one local maximum for $V(R)$, there is subset of the parameter space in which there are no forbidden regions for the shell. On the other hand, if $\Lambda<0$ then the potential would be {\bf monotonically increasing}, with a single root (which is the maximum radius), so the final outcome of the dynamics would always be a {\bf collapse}. 

\end{itemize}

\subsubsection{Equal masses at both sides, $\Lambda=0$ and uncharged}

\begin{itemize}

\item $-1\leq \omega <-(n-1)/n$

In this range there would be a local maximum for the potential, there is subset of the parameter space in which there are no forbidden regions for the shell.

\item $\omega =-(n-1)/n$

In this case the potential would be {\bf monotonically increasing}, if $2n<\kappa_DC$ with a single root (which is the maximum radius), so the final outcome of the dynamics would always be a {\bf collapse}. On the other hand, if $2n>\kappa_DC$ the shell can either expand indefinitely or collapse.

\item $-(n-1)/n <\omega \leq 1$

In this range the potential would be {\bf monotonically increasing}, with a single root (which is the maximum radius), so the final outcome of the dynamics would always be a {\bf collapse}. 

\end{itemize}

\section{Conclusions and final comments}
\label{final}

In this paper we thoroughly analysed different possible dynamics of a thin shell made of arbitrary matter satisfying the dominant energy condition in a spherically symmetric bulk with electric charge and cosmological constant in arbitrary dimensions.    
%It can be thought as a compendium of general characteristics of the motion for different matter-energy. 
%In Section \ref{sphericalmotion} the asymptotic behavior of the potential is studied in as much generality as possible without assuming an explicit matter model: the qualitative aspects of the dynamics of the shell are addressed in terms of the limiting values (for small and large $R$) of $p/\rho$. 
The analysis made in Section \ref{sphericalmotion} is arguably the most general asymptotic analysis that one can perform without assuming a definite matter-energy model.
As explained there, if we assume, besides DEC, reasonable hypothesis regarding the asymptotic behaviour of the matter-energy models within the evolving $n$-sphere, namely that there is no hysteresis and that $\alpha(R)$ has both a large $R$ limit and a small $R$ limit, the main qualitative aspects of the motion can be inferred from these limiting values and the parameters of the setting. 
In particular, we can exhaustively characterise the cases where a boundless expansion or a collapse are possible.

%There are many pieces of information that one can extract from these developments, and the interest of them depend on what kind of phenomena one is looking to describe. For example, if one is interested 

%For example, one can determine that in the case of barotropic fluids there are only a number of cases in which the oscillatory or stable static motion is possible...

%Do the table here!!!

On the other hand, in Section \ref{barotropic} we defined a family of equations of state, i.e. linear barotropic fluids, and this prescription allows us to be more specific in the description of the possible dynamics.  
However, the main difficulty in analysing the qualitative aspects of the different effective potentials lies in the fact that the signs of them and their derivatives depend in general on non-integral degree expressions whose roots and extremal points are complicated to address in full generality. Nevertheless, it applies the so-called ``Descartes' rule of signs'' which sets an upper bound on the number of positive roots for 
%both 
the effective potential: 
%and it first derivative: 
if we order the coefficients of the terms of the potential in descending order in $R$ then the number of sign changes in this sequence is the aforementioned upper bound. As discussed, the effective potential in full generality has eight terms, two of them always positive, another two can have either sign and the remaining four terms are always negative. In some situations, these four negative terms can be intertwined into the other four, so in principle there can be up to seven positive roots for the potential. However, it is straightforward to see in the potential (\ref{potential}) that the net contribution of the three terms that come from the quantity inside parenthesis that is being squared must be always negative. Then for the rule of signs these three terms should be taken as a single negative term whose order is given by the mass difference term (the higher order term), so it is as if they were six terms: three negative, two positive and one that can take either sign, which implies that there is at most five positive roots. In any case, this rule of thumb illustrates how complicated an analytic description of the potential can be. 

However, there are a number of cases where the description of the motion can be made simple. For example, the effective potential can acquire the form of a second order polynomial in $R^{-(n-1)}$, and the possible qualitative features can be fully described, as illustrated in Appendix \ref{quadratic}. In other cases, some of them included in the previous category, the potential is monotonic, either increasing or decreasing, and also in those cases the qualitative description of the motion can be fully addressed. But, in general, the expression that determine the qualitative features of the potential are of order higher than two and there is no monotonicity. Nevertheless, the cases described in Appendix \ref{quadratic} are useful for general situations as well, with variations that are related to the eventual presence of more than two roots in the potential. This is because, in qualitative terms, those cases are all the types of motion that the system can have: static (whether stable or unstable), oscillatory, collapsing, boundlessly expanding, or asymptotically approaching a finite radius. For example, if there are three roots for the potential then a combination of cases E, B and C is possible, each taking place at different ranges for $R$.
%which means that there are three types of dynamics taking place at different ranges of $R$: collapsing solutions, oscillatory solutions and boundlessly expanding solutions.
We are not specifying here what combinations take place to which $\omega$ and parameter space, as that would be extremely complicated and long to address. Instead, what we did in Section \ref{barotropic} is a general commentary for each subcase without having to analyse the potential in full generality, and a detailed description if the potential is a second order polynomial in $R^{-(n-1)}$ or a monotonic function.

\subsection{Weak Cosmic Censorship}

Regarding the weak cosmic censorship, a general analysis of the possibility of a gravitational collapse, as usually understood, can only partially be made looking at the results of the Subsection \ref{smallr}. Through that analysis we can characterise all the cases in which the possibility of a collision with the curvature sigularity at $r=0$ exists. Nevertheless, there are collapse solutions in which the shell rebounds into another asymptotically flat (de Sitter or Anti de Sitter) region through a wormhole, as illustrated for example in Section 4 of \cite{GaoLemos} for the Reissner-Nordstr\"om case ($\Lambda=0$) with a dust shell, and it never hits the curvature singularity. This can be understood by considering the corresponding maximal analytically extended manifold. In terms of the effective potential, these collapsing shells are characterised as bouncing solutions where the point of return represents a minimum radius for the shell and lies beyond the Cauchy horizon. 

%This situation represents a collapse because it makes the curvature singularity a naked one: the casual structure of the exterior solution is just like Minkowski spacetime, and it makes the $r=0$ singularities to have causal contact with future null infinity.

On the other hand, a solution that represents a shell colliding with the curvature singularity and with parameters that imply a naked singularity as the final state does not necessarily violate cosmic censorship. This is because the dynamics might not allow an initial configuration where both the shell lies in the domain of outer communications and no naked singularities are present beforehand. We can always choose both bulk solutions with parameters that make the $r=0$ singularity naked and find a solution $R(\tau)$ to the corresponding effective potential. And, as shown in the subsection \ref{smallr}, by choosing an appropriate $\alpha_0$ we ensure the existence of a solution that represents a shell reaching $r=0$. The problem is that this solution might not represent a shell with the standard orientation (it might glue two interior regions) or it might be defined only within the Cauchy horizon, that is, there might be a maximum radius for the shell smaller than the first root of at least one of the $F_i(R)$.       

In this way, we should narrow our discussion to shells with standard orientation (so we are allowed to label with $I$ the interior solution and with $II$ the exterior one), where the interior region is a black hole solution, that are initially defined outside the event horizon of this black hole. Then the weak cosmic censorship in this context is equivalent to saying that {\it if the exterior bulk solution implies a naked singularity at $r=0$ then it can not collapse}. And, as explained above, collapse here means that the shell crosses the event horizon of the inner black hole.  

Then, in this precise sense, as shown in Appendix \ref{censorship}, we can confidently state that {\bf Weak Cosmic Censorship hold for the entire class of solutions here described}. Although this might be the most important result of the present paper, the proof of it is given in an Appendix because it is closely related to Appendix \ref{specification}. This result generalises the analysis done in Section 3 of \cite{GaoLemos} and the one done in Subsection 5.4 of \cite{CrisostomoOlea} while being connected with them. It also complies with the results obtained in \cite{RochaSantarelli}, where the authors considered thin shells made of linear barotropic fluids in  a rotating odd-dimensional spacetime where all independent angular momenta coincide, for the spherically symmetric (zero angular momentum) case.

\section{Acknowledgements}

We thank Ernesto Eiroa for reading the article and useful comments. We also acknowledge Jorge Rocha for interesting and relevant questions and comments. Besides the institutions mentioned in the affiliations, we are grateful to the Instituto de Astronom\'{i}a Te\'orica y Experimental (IATE, CONICET - Universidad Nacional de C\'ordoba) for the hospitality. MAR and DA are supported by CONICET.

\appendix

\section{\bf Proof of the consistency of the choice of $\xi_i$}
\label{specification}

Here we show that the specifications (\ref{espureoI},\ref{espureoII}) are consistent in the sense that the expressions within those specifications can only have a root (at $R_{\xi_i}$) in a forbidden region ($V(R_{\xi_i})>0$) or inside (on) an event horizon ($F_i(R_{\xi_i})\leq 0$). In the last case, a change of sign of $\partial r/\partial \eta$ would not imply an inconsistency with the timelike nature of the shell as the $r$ coordinate would be a time (null) coordinate. A simple substitution leads us to the expression
\begin{equation}
\label{conditionespureo}
\left.\left(\frac{\partial r}{\partial \eta}\right|_{\eta=0^{i}}\right)^2=\dot{R}^2+F_i=-V(R)+F_i(R)=\left[\frac{n(F_i-F_j)}{2\kappa_D\rho R}+\frac{\kappa_D \rho R}{2n}\right]^2\geq 0,
\end{equation}  
where $j$ is the index different from $i$ and the roots of $\xi_i(R)$ are the same as the roots of the right hand side of the above equation. 
%APPARENTLY FOR II WE CHOOSE THE EXPRESSION INSIDE BRACKETS WITH THE OPPOSITE SIGN !
%The roots of $\xi_i$ are then solutions of the equation
%\begin{equation}
%\label{root}
%n^2|F_I(R_{\xi_i})-F_{II}(R_{\xi_i})|=\kappa_D^2\rho^2(R_{\xi_i})R^2_{\xi_i},
%\end{equation}
%and each root of (\ref{root}) is a root of either $\xi_I$ or $\xi_{II}$: if $F_I(R_{\xi_i})>F_{II}(R_{\xi_i})$ then $i=II$, while if $F_{II}(R_{\xi_i})>F_I(R_{\xi_i})$ then $i=I$.
From (\ref{conditionespureo}) one can see that if $V(R_{\xi_i})\leq 0$ then $F_i(R_{\xi_{i}})=V(R_{\xi_{i}})\leq 0$, so if the locus $r=R_{\xi_i}$ is not in a forbidden region then it is within (on) an horizon. %On the other hand, at $R_{\xi_i}$ we have
%\begin{equation}
%V(R_{\xi_i})=\frac{F_I(R_{\xi_i})+F_{II}(R_{\xi_i})}{2}-\frac{|F_I(R_{\xi_i})-F_{II}(R_{\xi_i})|}{2}
%\end{equation}
%so if we assume that both $F_i(R_{\xi_i})>0$, then $|F_I-F_{II}|<F_I+F_{II}$, which in turn implies $V(R_{\xi_i})>0$. In this way, if $R_{\xi_i}$ is not within an horizon, then it is in a forbidden region.     

On the other hand, it is not obvious whether the solutions of (\ref{eqnmotion1}) are solutions of (\ref{Israelndima}) at the radius $R_{\xi_i}$ because of the ill-defined value of the corresponding extrinsic curvature (\ref{extrinsic2}) thereat. Nevertheless, one can rewrite the extrinsic curvature to show that the singularity is only apparent, stemming from the ill-defined derivative of the function $x^{1/2}$ at $x=0$. The specifications (\ref{espureoI},\ref{espureoII}) are chosen such that (\ref{conditionespureo}) implies the following
\begin{eqnarray}
\left.\frac{\partial r}{\partial \eta}\right|_{\eta=0^{-}} &= & \frac{n(F_I-F_{II})}{2\kappa_D\rho R}+\frac{\kappa_D \rho R}{2n} \\
\left.\frac{\partial r}{\partial \eta}\right|_{\eta=0^{+}} &= & \frac{n(F_I-F_{II})}{2\kappa_D\rho R}-\frac{\kappa_D \rho R}{2n}. 
\end{eqnarray}
In this way, the potentially divergent component of the extrinsic curvature can be written as
\begin{eqnarray}
K^{\; \tau}_{I \; \tau} =\left. \frac{1}{2} \frac{\partial f}{\partial \eta} \right|_{\eta=0^-} &= F_{I}^{-1}(R)\dot{R}\frac{\partial}{\partial \tau}\left.\left(\frac{\partial r}{\partial \eta}\right|_{\eta=0^-}\right)= &F_I^{-1}(R) \dot{R}^2 \frac{d}{dR} \left(\frac{n(F_I-F_{II})}{2\kappa_D\rho R}+\frac{\kappa_D \rho R}{2n}\right) \\
K^{\; \tau}_{II \; \tau} =\left. \frac{1}{2} \frac{\partial f}{\partial \eta} \right|_{\eta=0^+} &= F_{II}^{-1}(R)\dot{R}\frac{\partial}{\partial \tau}\left.\left(\frac{\partial r}{\partial \eta}\right|_{\eta=0^+}\right)= &F_{II}^{-1}(R) \dot{R}^2 \frac{d}{dR} \left(\frac{n(F_I-F_{II})}{2\kappa_D\rho R}-\frac{\kappa_D \rho R}{2n}\right),
\end{eqnarray}  
where it is clear that the extrinsic curvature at both sides do not have a singular point at any of the $R_{\xi_i}$ roots, provided they exist\footnote{Even in the very special case where the root $R_{\xi_i}$ lies precisely at the corresponding horizon, the coefficient $\dot{R}^2/F_{i}$ at that point would acquire the value $-1$ as it can be seen from (\ref{conditionespureo}).}.

\section{\bf Proof of Weak Cosmic Censorship for this class of solutions.}

\label{censorship}

First we recall 
\begin{equation}
F_i(R)=R^{-(2(n-1))}\left(-2\frac{\Lambda}{n(n+1)} R^{2n}+R^{2(n-1)}-2M_iR^{n-1}+Q_i^2\right),
\end{equation}
which is positive at sufficiently small $R$ for non-zero $Q_i$ (negative in the uncharged case) and it has the opposite sign to $\Lambda$ for large $R$ (being positive in the asymptotically flat case). Because of Descartes' rule of signs, there can be up to three positive roots for this expression, which is the case of a subextremal Reissner-Nordstr\"om-de Sitter solution. But there might be as well two, one or zero positive roots depending on the coefficients, and the causal structure of the corresponding maximal analytically extended solution strongly depends on them.

We assume that the shell has the standard orientation, and consequently label $I$ the interior region and $II$ the exterior one. This assumption implies that in some region where both $F_i>0$ we must have $\xi_{II}=+1$, so
\begin{equation}
\label{standardcondition}
F_I(R)>\frac{\kappa_D^2\rho^2R^2}{n^2}+F_{II}(R).
\end{equation}
As discussed in Section \ref{sphericalshells}, this inequality implies $\xi_I=+1$ as well. We also assume that the interior spacetime contains a black hole, so $F_I(R)$ has at least one root $R_{hI}$ where $F_I'(R_{hI})>0$, and that $F_{II}(R)>0$ in its entire range, so if the shell reaches $r=0$ then a naked singularity would appear. In this way, (\ref{standardcondition}) can not hold for all $R>R_{hI}$, because $F_I(R_{hI})=0$ and the right hand side of the inequality is positive for all $R$. Then $\xi_{II}(R_{hI})=-1$, so there must be a root $R_{\xi_{II}}>R_{hI}$ of $\xi_{II}(R)$ where (\ref{standardcondition}) turns into an equality, and, as shown in the previous Appendix, this root must lie in a forbidden region ($V(R_{\xi_{II}})>0$). In this way, provided there is an initial configuration in the region outside the horizon where the shell has the standard orientation, that is $R_0>R_{\xi_{II}}>R_{hI}$, there must be a point of return $V(R_c)=0$ such that $R_{hI}<R_{\xi_{II}}<R_c<R_0$. This implies that $R_c$ represents the minimum radius for the kind of solutions we are considering, so the shell never crosses the horizon and {\bf Weak Cosmic Censorship holds}. 

% In this way, provided there is an initial configuration $V(R_0)\leq 0$ such that both $F_{i}(R_0)>0$ then 

%\section{\bf Proof of existence of Two-world Orbits in the standard oriented cases that do not collapse into the singularity}

%\label{oscillations}

%If a shell has a standard orientation and crosses the event horizon, then inequality (\ref{standardcondition}) proves that this event horizon must correspond to region $II$, that is it must lie at $R=R_{hII}$, where $F_{II}(R_{hII})=0$, and if there is also an ``interior horizon'' $R_{hI}$ then $R_{hII}>R_{hI}$. On the other hand, (\ref{conditionespureo}) implies that if $F_i(R)\leq 0$ for any $i$ then $V(R)\geq 0$, that is, those ranges for $R$ within event horizons are always permitted regions. In this way, a collapsing shell with the standard orientation  can not have a point of return in a region with $F_i(R)<0$, so if it crosses the horizon there are two possibilities: it hits the curvature singularity or it bounces in a black hole region where $F_i(R)>0$. Those are the so-called Two-world Orbits.  

\section{\bf Criteria to determine the qualitative aspects of the dynamics when the potential is a second order polynomial in $R^{-(n-1)}$}
\label{quadratic}

In Section \ref{barotropic} there are a number of cases in which the potential turns out to be a second order polynomial in $R^{-(n-1)}$. When this happens all the qualitative aspects of the motion can be easily derived in terms of the coefficients. If the potential acquires the following form
\begin{equation}
V(R)=a+bR^{-(n-1)}+cR^{-2(n-1)},
\end{equation}
then a simple change of variable $u=R^{-(n-1)}$ reveals the possible qualitative aspects of the motion. $V(u)$ is a second order polynomial, restricted to $u>0$, and the local extremal points $u_m$ of $V(u)$ must be local extremal points $R_m$ of $V(R)$, where $R_m=u_m^{-1/(n-1)}$. In this way, we can derive all the relevant qualitative aspects simply by analysing the sign of the discriminant $\Delta=b^2-4ac$ and the signs of each coefficient. In terms of the signs of $a$, $b$, $c$ and $\Delta$ the following cases are possible:

\begin{figure}[h]
\label{quadraticcases}
\centerline{\includegraphics[width=.75\textwidth]{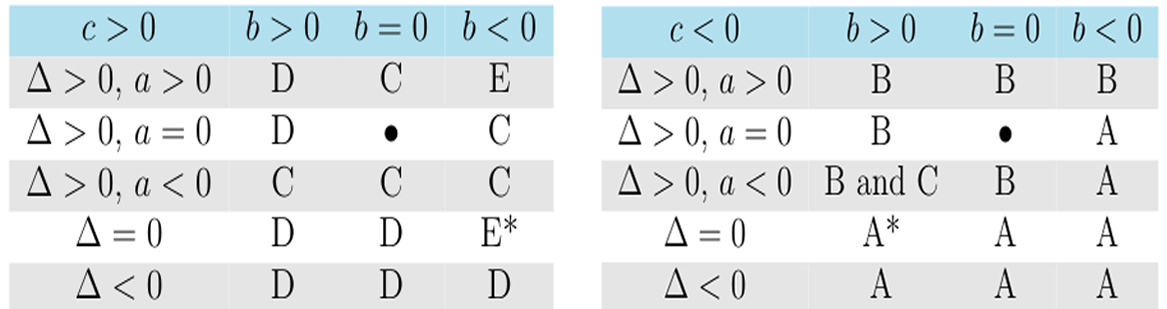}}
\end{figure}
where the capital letters represent the following dynamical behaviours:

\begin{itemize}
\item $A$: there are no forbidden regions. The shell will either expand indefinitely or collapse depending on the sign of the initial velocity.
\item $A^*$: it is a special subcase of the previous one, as there are no forbidden regions as well, but an unstable static solution is possible. Apart from this, the shell will either expand indefinitely, collapse, or asymptotically approach the radius of the static solution either from above or below.  
\item $B$: there is a maximum radius. This means that if the shell is initially expanding it would reach a maximum radius and collapse afterwards. The final state of the evolution is always a collapse.
\item $C$: there is a minimum radius. This means that if the shell is initially contracting it would rebound at a certain radius and expand indefinitely afterwards. The final state of the evolution is always an indefinite expansion.
\item $B$ and $C$: both classes of solutions coexist, which means that the minimum radius of the expanding solutions is greater than the maximum radius of the collapsing ones.
\item $D$: there is no solution.
\item $E$: there are only oscillating solutions. The motion is bounded between a minimum and a maximum radius, and it is periodic. 
\item $E^*$: there is only one possible solution: a stable static one.
\end{itemize}

\section{\textbf{Vlasov matter}}

Vlasov matter is another relevant and simple matter model used for different astrophysical settings \cite{VlasovReview}. 
%It has the advantage that its range of applicability is phenomenologically clear:
It is a reasonable model for a physical system if it
% can be thought of as an Einstein-Vlasov system whenever it 
is composed of many constituents with low collision probability among themselves. If this system is also self-gravitating then it can be mathematically described as an Einstein-Vlasov system: an ensemble of collisionless particles that interact with each other only by means of the curvature that they generate as a whole. As a consequence of this setting, the constituent particles must follow geodesic trajectories. In the context of thin shells it is not obvious what a ``geodesic'' is because of the discontinuity of the metric connection on the shell. Furthermore, in general there is no curve contained within the evolving shell that is a geodesic of any of the metric connections defined there (see \cite{AndersonTavakol2} and \cite{SeahraWesson} for a discussion in the context of brane cosmology). One then typically imposes that \textit{the particles follow geodesics of the induced metric}, which is particularly reasonable in the spherically symmetric case as such geodesics are simply the trajectories of conserved angular momentum within the shell. The system we are considering in this Appendix is a $D$-dimensional generalization of the four-dimensional case already analysed in \cite{paper1} and references therein.

%Ecuaci�n de Vlasov?

Assuming that all particles are identical, the $S$ tensor and the particle density current can be written as
\begin{equation}
S^{ij} =- \mu \int f(x,p) \sqrt{-h} u^i u^j \frac{dp^1..dp^n}{p_0} \;\;\; , \;\;\; N^i=\int f(x,p) \sqrt{-h} u^i \frac{dp^1..dp^n}{p_0},
\end{equation}
where $\mu$ is the particles proper mass and $f(x,p)$ is the distribution of the number of particles in the tangent bundle. Taking into account the symmetry, the distribution of the number of particles with angular momentum modulus $L$ (which we call $n(L)$) must be conserved, and the independent components of the $S$ tensor can be written in terms of that function as follows
\begin{equation}
\label{rhopvlasov}
\rho(R) = \frac{\mu}{S_nR^{n+1}} \int n(L) \sqrt{R^2+L^2} dL \;\;\; , \;\;\; p(R)=\frac{\mu}{nS_nR^{n+1}} \int \frac{n(L)L^2}{\sqrt{R^2+L^2}} dL,
\end{equation}
where $S_n$ is the area of a $n$-sphere of unit radius.

We define
\begin{equation}
f(R) \equiv \int n(L) \sqrt{R^2+L^2} dL.
\end{equation}
For the sake of simplicity we will assume an uncharged and asymptotically flat setting, so the equation of motion for the shell can be written as follows
\begin{equation}
\label{eqnmotionV}
V(R)= \frac {1}{2} - \frac{M_I + M_{II}}{2R^{n-1}} -\frac{n^2(M_{II}-M_{I})^2R^2}{2C_n^2 f(R)^2}-\frac{C_n^2 f(R)^2}{8n^2R^{2n}},
\end{equation}
where $C_n \equiv \kappa \mu/S_n$.

If the function $n(L)$ has compact support, then $\alpha_{\infty}=0$, $\alpha_0=1/n$, and the asymptotic behavior of $V(R)$ can be described as follows
\begin{eqnarray}
 R\to 0 & \;\;\;\;\;\; & V(R) \to -\frac{C_n^2N^2\langle L \rangle^2}{8n^2R^{2n}} \\
R\to \infty & \;\;\;\;\;\; & V(R) \to \frac{C_n^2N^2- n^2(M_{II}-M_I)^2}{2C_n^2N^2},
\end{eqnarray}
where $\langle L \rangle$ is the mean angular momentum modulus. One can notice that \textit{the shell can have unbounded motion if and only if $C_n N <n (M_{II}-M_I)$}, which coincides with the analysis made in Subsection \ref{simplecase1}, and that there are always collapsing solutions for a small enough initial radius, in concordance with Subsection \ref{simplecase2}.
Furthermore, taking derivatives of (\ref{eqnmotionV}), it can be shown that \textit{there are oscillating or static solutions only if $n=2$}, which is the case analysed in \cite{paper1}.
%It can also be shown that (\ref{espureo}) implies that region $II$ must be an {\it exterior} Schwarzschild region for both the unbounded and the oscillating solutions.


\begin{thebibliography}{99}
\bibitem{Israel} Israel W, \textit{Nuov. Cim.} B \textbf{44}, 1 (1966).
%\bibitem{RS1} Randall L, Sundrum R, \textit{Phys. Rev. Lett.} {\bf 83}, 4690-4693 (1999).
%\bibitem{RS2} Randall L, Sundrum R, \textit{Phys. Rev. Lett.} {\bf 83}, 3370-3373 (1999).
\bibitem{Review} Maartens R, Koyama K, ``Brane-World Gravity'' \textit{Living Rev. Rel.} {\bf 13}, 5 (2010) \textit{http://www.livingreviews.org/lrr-2010-5}.
%\bibitem{emission}  
\bibitem{censor} Barrabes C, Israel W, Letelier P S, \textit{Phys. Lett. A} {\bf 160}, 41-44 (1991).
\bibitem{astrophysics} Yangurazova L R, Bisnovatyi-Kogan G S, \textit{Astroph. and Space Sci.} {\bf 100}, 319-328 (1984).
\bibitem{phasetransition} Berezin V A, Kuzmin V A, Tkavech I I, \textit{Phys. Rev. D} {\bf 36}, 10 2919 (1987) \\
Ansoldi S, Guendelman E, Shilon S, \textit{Proceedings of BH2: Dynamics and Thermodynamics of Black Holes and Naked Singularities} - Politecnico di Milano, Italy (2007).
\bibitem{paper1} Gleiser R J, Ramirez M A, \textit{Class. Quant. Grav.} \textbf{26}, 045006 (2009).
%\bibitem{DafermosRendall} Dafermos M, Rendall A D, {\it Ann. Henri Poincare} {\bf 6}, 1137-1155 (2005).
\bibitem{EiroaSimeone} Eiroa E F, Simeone C, {\it Int. J. Mod. Phys. D} {\bf 21}, 125003301 (2012).
\bibitem{CrisostomoOlea} Cris\'ostomo J, Olea R, {\it Phys. Rev. D}, {\bf 69}, 104023 (2004).
\bibitem{TSE} Thibeault M, Simeone C, Eiroa E F, {\it Gen. Relativ. Grav.} {\bf 38}, 1593 (2006).
\bibitem{Ref9ES} Richarte M G, Simeone C, {\it Phys. Rev. D}, {\bf 76}, 087502 (2007); {\bf 77}, 089903(E) (2008); Habib Mazharimousavi S, Halilsoy M, Amirabi Z, {\it Phys. Rev. D} {\bf 81}, 104002 (2010);
Simeone C, {\it Phys. Rev. D} {\bf 83}, 087503 (2011).
\bibitem{GaoLemos} Gao S, Lemos J P S, {\it Int. J. Mod. Phys. A} {\bf 23}, 2943 (2008).
\bibitem{BJB} Banerjee A, Jusufi K, Bahamonde S, {\it Grav. Cosmol.} {\bf 24}, 1 71-79 (2018). 
\bibitem{tangherlini} Tangherlini F R, {\it Nuov. Cim.} {\bf 27}, 636-651 (1963).
\bibitem{myersperry} Myers R C, Perry M J, {\it Ann.
Phys.} (NY) {\bf 172}, 304 (1986).

%\bibitem{Z2} Lukas A, Ovrut B A, Stelle K S, Waldram D, {\it Phys. Rev. D} {\bf 59}, 086001 (1999).
%\bibitem{SMS} Shiromizu T, Maeda K, Sasaki M, {\it Phys. Rev. D} {\bf 62} 024012, 1-6 (2000).
%\bibitem{BCG} Bowcock P, Charmousis C, Gregory R, \textit{Class. Quant. Grav.} {\bf 17}, 4745-4763 (2000).
\bibitem{FST} Fayos F, Senovilla J M M, Torres R, {\it Phys. Rev. D} {\bf 54}, 4862-4872 (1996).
\bibitem{GoldwirthKatz} Goldwirth D S, Katz J, {\it Class. Quant. Grav.}{\bf 12}, 769-777 (1995).
%\bibitem{Stoyca} Stoyca H, Tye S-H H, Wasserman I, \textit{Phys Lett. B} {\bf 482}, 205-212 (2000).
%\bibitem{DavisDavis} Davis A-C, Davis S C, Perkins W B, Vernon I R, \textit{Phys. Lett. B} {\bf 504}, 254 (2001).
\bibitem{RochaSantarelli} Rocha J V, Santarelli R, {\it Class. Quantum Grav.} {\bf 35}, 125009 (2018).
\bibitem{VlasovReview} Andr\'easson H, ``The Einstein-Vlasov system/ Kinetic theory'' \textit{Living Rev. Rel.}
\textbf{14}, 4 (2011).
%\bibitem{LangloisMaeda} Langlois D, Maeda K-I, Wands D, {\it Phys. Rev. Lett.} {\bf 88}, 181301 (2002).
%\bibitem{GravanisWillison} Gravanis E, Willison S, {\it J. Math. Phys.} {\bf 47}, 092503 (2006) \\ {\it J. Geom. Phys.} {\bf 57}, 1861-1882 (2007).
\bibitem{AndersonTavakol2} Anderson E, Tavakol R, {\it JCAP} {\bf 0510}, 017 (2005).
\bibitem{SeahraWesson}Seahra S S, Wesson P S, {\it Class. Quant. Grav.} {\bf 20}, 1321-1340 (2003).
%\bibitem{LCCGVP} Long J C, Chan H W, Churnside A B, Gulbis E A, Varney M C M, Price J C, {\it Nature} {\bf 421}, 922-925 (2003).

%\bibitem{WMAP} Komatsu E, \textit{et al.}, \textit{Astrophys. J. Suppl. Ser.} {\bf 192}, 18 (2011).
%\bibitem{DahiaRomero} Dahia F, Romero C, da Silva L F P, Tavakol R, \textit{Gen. Rel. Grav.} {\bf 40}, 1341-1351 (2008).
%\bibitem{GuhaChakraborty} Guha S, Chakraborty S, {\it Gen. Rel. Grav.} {\bf 42}, 1739-1754 (2010).
%\bibitem{Christodoulou} Christodoulou D, ``The formation of shocks in 3-dimensional fluids'', {\it European Mathematical Society} (2007).
%\bibitem{paper2} Gleiser R J, Ramirez M A, \textit{Class. Quant. Grav.} \textbf{27}, 065008 (2010).
\end{thebibliography}
\end{document}